%% file: main.tex
\documentclass[11pt,letterpaper]{article}
\input{preamble}

\input{notation}

\title{A Quantum Lower Bound for Triangle Finding}
\author{%
  Arjan Cornelissen\thanks{Department of Computer Science and Technology,
    University of Cambridge. \texttt{ajcornelissen@outlook.com}.}
  \qquad
  Amin Shiraz Gilani\thanks{Joint Center for Quantum Information and Computer
    Science, University of Maryland. \texttt{asgilani@umd.edu}.}
  \qquad
  Xingyu Zhou\thanks{Department of Computer Science, The University of British
    Columbia. \texttt{zxingyu@cs.ubc.ca}.}}
\date{}
\hypersetup{
  pdftitle={A Quantum Lower Bound for Triangle Finding},
  pdfauthor={Arjan Cornelissen, Amin Shiraz Gilani, and Xingyu Zhou},
  pdfkeywords={quantum query complexity, triangle finding, lower bounds}
}

\begin{document}
\maketitle
\thispagestyle{empty}
\input{sections/abstract}

\clearpage
\tableofcontents
\clearpage

\input{sections/introduction}
\input{sections/overview}
\input{sections/preliminaries}
\input{sections/tools}

\input{sections/triangle}

\section*{Acknowledgments}
We thank Aleksandrs Belovs for suggesting that his framework based on
a hidden type could be applied to lower bounds for triangle finding. We also thank Subhasree Patro, Zeph Landau and Thiago Bergamaschi for motivating discussions.
AC is supported by ERC Starting Grant 101163189. ASG received support from the National Science Foundation (grant 26-17356). XZ is supported by NSERC Grant RGPIN-2024-06493.

\paragraph{Use of generative AI.}
Our initial proof established an \(\Omega(n\sqrt{\log n})\) lower bound by building on Belovs’s approach with additional constructions and estimates with the help of OpenAI GPT 6 Astra. Starting from this proof, GPT 6 Astra developed the entire argument yielding the improved \(\Omega(n^{13/12})\) bound.

OpenAI GPT 6 Astra, and Anthropic Opus 5.5 were used for literature searches, manuscript drafting and revision, proof development and checking, and LaTeX preparation. The authors retain full responsibility for reviewing the generated material and for all mathematical claims and final text.

\begingroup
\small
\raggedright
\bibliographystyle{alphaurl}
\AddToHookNext{cmd/thebibliography/after}{%
  \addcontentsline{toc}{section}{References}%
}
\bibliography{references}
\endgroup
\end{document}

%% file: preamble.tex
\usepackage[T1]{fontenc}
\usepackage{lmodern}
\usepackage[letterpaper,margin=1in]{geometry}
\usepackage{amsmath,amssymb,amsthm,mathtools,mathrsfs}
\usepackage{microtype}
\usepackage{xcolor}
\usepackage{graphicx}
\usepackage{tikz}
\usepackage{booktabs,array}
\usepackage{flafter}
\usepackage{enumitem}
\usepackage{hyperref}
\hypersetup{
  colorlinks=true,
  linkcolor=blue,
  citecolor=blue,
  urlcolor=blue,
  filecolor=blue,
  linkbordercolor={0 0 1},
  citebordercolor={0 0 1},
  urlbordercolor={0 0 1},
  filebordercolor={0 0 1}
}
\AtBeginDocument{\hypersetup{pdfborder={0 0 0.7}}}
\usepackage{aliascnt}
\newtheorem{theorem}{Theorem}[section]
\newaliascnt{lemma}{theorem}
\newtheorem{lemma}[lemma]{Lemma}
\aliascntresetthe{lemma}
\newaliascnt{proposition}{theorem}

\aliascntresetthe{proposition}
\newaliascnt{corollary}{theorem}
\newtheorem{corollary}[corollary]{Corollary}
\aliascntresetthe{corollary}
\newtheorem*{nonbipartitecorollary}{Corollary~\ref{cor:non-bipartite}}
\theoremstyle{definition}
\newaliascnt{definition}{theorem}

\aliascntresetthe{definition}
\theoremstyle{remark}
\newaliascnt{remark}{theorem}

\aliascntresetthe{remark}
\usepackage[capitalise,nameinlink,noabbrev]{cleveref}
\numberwithin{equation}{section}

%% file: notation.tex
\newcommand{\E}{\mathbb E}
\newcommand{\one}{\mathbf 1}
\newcommand{\Op}[1]{\mathsf{#1}}
\newcommand{\identity}{\Op{I}}
\newcommand{\norm}[1]{\lVert #1\rVert}

\newcommand{\ip}[2]{\langle #1,#2\rangle}
\newcommand{\ket}[1]{\lvert #1\rangle}
\newcommand{\bra}[1]{\langle #1\rvert}
\newcommand{\braket}[2]{\langle #1\mid #2\rangle}
\DeclareMathOperator{\st}{st}
\DeclareMathOperator{\supp}{supp}
\DeclareMathOperator{\Bernoulli}{Bernoulli}
\newcommand{\vx}{\boldsymbol{x}}
\newcommand{\vy}{\boldsymbol{y}}
\newcommand{\vz}{\boldsymbol{z}}
\newcommand{\vzero}{\boldsymbol{0}}
\newcommand{\vone}{\boldsymbol{1}}
\newcommand{\vX}{\boldsymbol{X}}
\newcommand{\vY}{\boldsymbol{Y}}
\newcommand{\vu}{\boldsymbol{u}}
\newcommand{\va}{\boldsymbol{a}}
\newcommand{\vf}{\boldsymbol{f}}
\newcommand{\vg}{\boldsymbol{g}}
\newcommand{\vh}{\boldsymbol{h}}
\newcommand{\Triangles}{\mathcal T}
\newcommand{\Certificates}{\mathcal F}
\newcommand{\W}{\mathscr W}
\newcommand{\Wrest}{\mathscr W_{\mathrm{rest}}}
\newcommand{\Input}{\mathscr H}
\newcommand{\X}{\mathscr X}
\newcommand{\Y}{\mathscr Y}
\newcommand{\Cond}{\Op{E}}
\newcommand{\Deriv}{\Op{D}}
\newcommand{\Phase}{\Op{Q}}
\newcommand{\Oracle}{\Op{O}}
\newcommand{\Unitary}{\Op{U}}
\newcommand{\Comm}{\Op{K}}
\newcommand{\Piacc}{\Pi_{\mathrm{acc}}}
\newcommand{\Star}{\Op{L}}
\newcommand{\Energy}{\Op{H}}
\newcommand{\Filter}{\Op{F}}
\newcommand{\Avg}{\Op{A}}
\newcommand{\Local}{\Op{T}}
\newcommand{\Ups}{\Upsilon}

%% file: sections/abstract.tex
\begin{abstract}
Triangle finding is a fundamental problem in graph theory that has played a central role in the development of quantum algorithmic techniques. For $N$-vertex graphs given by their adjacency matrices, a long line of work has improved the quantum query upper bound from $O(N^{3/2})$ to $O(N^{5/4})$. However, the best known lower bound has remained the elementary $\Omega(N)$ bound obtained from unstructured search. We prove the first superlinear quantum query lower bound for triangle finding, showing that $\Omega(N^{13/12})$ queries are necessary. Our result also implies the same lower bound for triangle detection and for detecting every fixed connected non-bipartite graph.

The proof builds on the compressed oracle approach of Zhandry and the transfer approaches of Belovs. It compares the computation on graphs with a planted triangle to the computation on random graphs conditioned to be triangle-free, through a map on the input registers that almost commutes with every query. We average over local changes to the input while preserving its distribution, through a procedure known as \emph{conditional resampling}. Our main technical contribution is a set of estimates that let us control the query error under this averaging, despite the dependence introduced by conditioning a dense graph to be triangle-free.
\end{abstract}

%% file: sections/introduction.tex
\section{Introduction}\label{sec:intro}

Triangle finding asks for three pairwise adjacent vertices in a graph given
by queries to its adjacency matrix. It is a basic test of how quantum
algorithms exploit the structure of a small certificate: a solution consists
of only three edges, but the possible solutions overlap in many ways.
For a graph on $N$ vertices, classical randomized algorithms require
$\Theta(N^2)$ queries in the worst case, whereas searching over all vertex
triples with Grover's algorithm uses $O(N^{3/2})$ quantum
queries~\cite{Gro96}. The elementary quantum lower bound is
$\Omega(N)$, obtained by embedding unstructured search~\cite{BBBV97}.

Triangle finding has helped drive the development of quantum algorithmic techniques, with a long line of work yielding progressively better
quantum query upper bounds. \cref{tab:triangle-history} summarizes this
progress. The best known quantum query upper bound exponent is
$5/4$~\cite{LG14,CLM20,Kin26}. In contrast, the lower bound
has remained linear. The difficulty of proving a superlinear bound has been
noted repeatedly in work on quantum graph
algorithms~\cite{MSS07,CK11,LG14,CGP26}.

\begin{table}[htbp]
 \centering
 \begin{tabular}{@{}lll@{}}
  \toprule
  Approach & Quantum queries & Reference \\
  \midrule
  Search over triples
   & $O(N^{3/2})$
   & Grover~\cite{Gro96} \\
  Combinatorial arguments and search
   & $\widetilde O(N^{10/7})$
   & Szegedy~\cite{Sze03} \\
  Quantum walks
   & $\widetilde O(N^{13/10})$
   & Magniez--Santha--Szegedy~\cite{MSS07} \\
  Learning graphs
   & $O(N^{35/27})$
   & Belovs~\cite{Bel12} \\
  Learning graphs
   & $O(N^{9/7})$
   & Lee--Magniez--Santha~\cite{LMS13} \\
  Nested quantum walks
   & $\widetilde O(N^{9/7})$
   & Jeffery--Kothari--Magniez~\cite{JKM13} \\
  Combinatorial arguments and walks
   & $\widetilde O(N^{5/4})$
   & Le Gall~\cite{LG14} \\
  Extended learning graphs (ELGs)
   & $O(N^{5/4})$
   & Carette--Lauri\`ere--Magniez~\cite{CLM20} \\
  Combinatorial arguments and ELGs
   & $N^{5/4}/2^{\Omega(\log^* N)}$
   & Kintali~\cite{Kin26} \\
  \bottomrule
 \end{tabular}
 \caption{Progress in quantum query upper bounds for triangle finding in
 the adjacency matrix model. The first row is the direct application of
 quantum search to vertex triples. The notation $\widetilde O$ suppresses
 polylogarithmic factors. In the last row, $\log^* N$ denotes the iterated binary logarithm.}
 \label{tab:triangle-history}
\end{table}

The small size of a triangle certificate is itself an obstacle for some
lower bound techniques. The certificate complexity barrier limits the
adversary method with nonnegative weights to $O(N)$ for triangle
detection~\cite{Zha04,SS06}. Gilani and Zhou~\cite{GZ26} recently proved
superlinear quantum query lower bounds for larger cliques and, more
generally, every fixed connected subgraph of chromatic number at least four.
Their applications do not yield a superlinear bound for triangles.

\subsection{Results}\label{sec:intro-approach}

In this work, we prove the first superlinear lower bound for the triangle
finding problem.

\begin{theorem}[Main result]
    \label{thm:main-result}
    The quantum query complexity of the triangle finding problem is $\Omega(N^{13/12})$, where $13/12\approx1.0833$.
\end{theorem}

\begin{figure}[htbp]
 \centering
 \input{figures/bound-timeline}
 \caption{Bounds on the exponent of the quantum query complexity of triangle
 finding in the adjacency matrix model. Subpolynomial factors are omitted;
 \cref{tab:triangle-history} gives the precise upper bounds. }
 \label{fig:bound-timeline}
\end{figure}
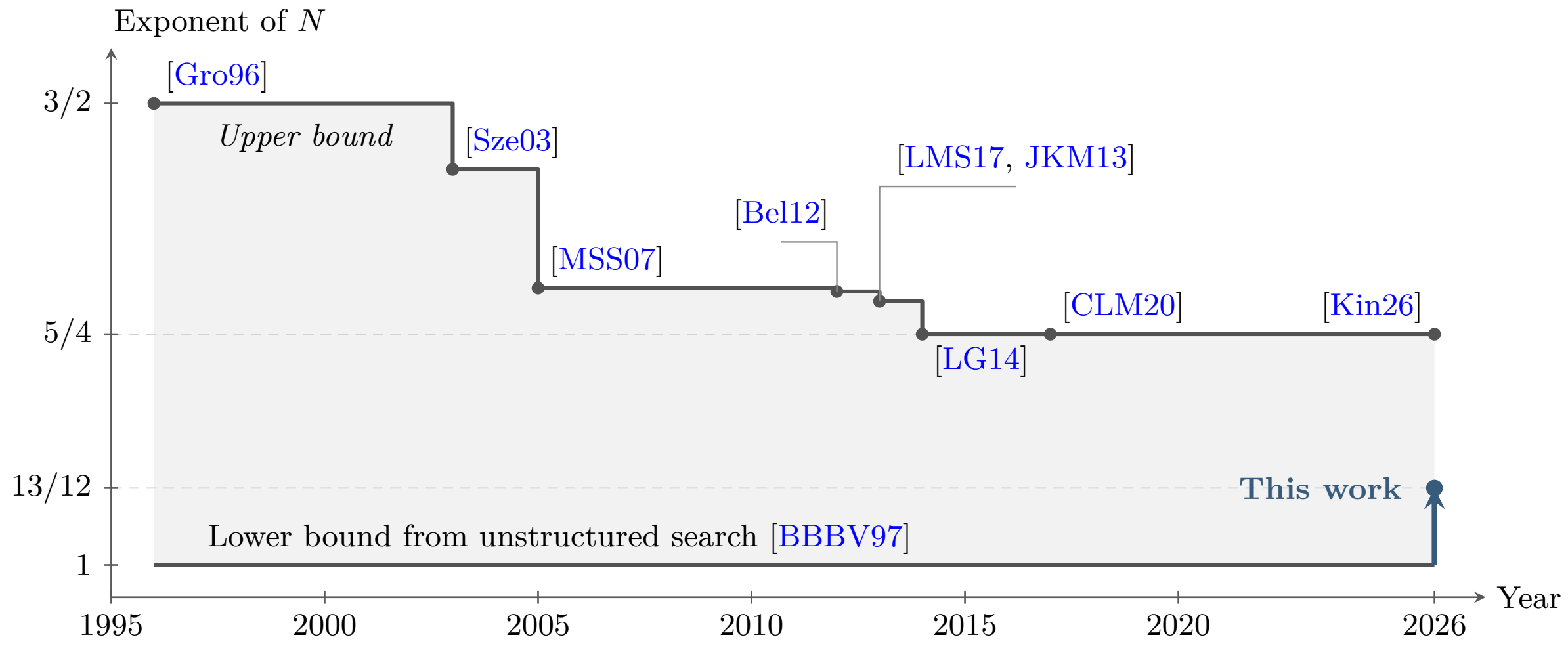

\Cref{fig:bound-timeline} places this lower bound alongside the preceding
upper bounds.

Combining the search-to-decision reduction for triangles~\cite[Lemma 23]{CGP26}
with the subgraph detection reduction of~\cite[Corollary 5.3]{GZ26} gives the
following consequence.

\begin{corollary}\label{cor:non-bipartite}
For every fixed connected non-bipartite graph $H$, detecting $H$ as a
subgraph of an $N$-vertex graph in the adjacency matrix model requires
$\Omega(N^{13/12})$ quantum queries.
\end{corollary}

We obtain \Cref{thm:main-result} using a novel framework for lower bounding quantum query complexity recently introduced by Belovs~\cite{Bel26}. The framework in fact shows a lower bound for a more general task. Fix any graph $G$ known to the
algorithm. Write $m\coloneqq|E(G)|$, let $\Delta$ be its maximum degree, and let
$M\ge1$ be its number of triangles. Choose a triangle of $G$ uniformly, set its
edges to one, and sample the other edges from the product $\Bernoulli(p)$ distribution conditioned on
that triangle being unique. Under this distribution, we have the following theorem.

\begin{theorem}[Informal version of \Cref{thm:fixed-graph}]\label{thm:main}
    There are absolute constants $C_0,C_1>0$, such that
    \[
     0<p\le\frac1{10\sqrt\Delta},\qquad
     \Delta p\ge C_0\log(2m)
     \quad\Longrightarrow\quad
     \sqrt\sigma\le C_1(t+3)\frac{\Delta^{1/3}}{\sqrt M\,p^{1/6}},
    \]
    where $\sigma$ is the average success probability of any $t$-query quantum algorithm that outputs the hidden triangle under this distribution.
\end{theorem}

For the complete tripartite graph $G=K_{n,n,n}$, we have $m=3n^2$, $\Delta=2n$, and $M=n^3$, giving the claimed bound in \cref{thm:main-result} with $p = \Theta(n^{-1/2})$. The same lower bound holds for unrestricted triangle detection, by the
constant-factor search-to-decision reduction for triangles~\cite[Lemma 23]{CGP26}.

The proof compares these positive inputs with Bernoulli graphs
conditioned to be triangle-free, at density $p=\Theta(n^{-1/2})$. At this density an
unconditioned Bernoulli graph has about $n^{3/2}$ triangles and is easy to search.
Conditioning removes all of them while keeping a simple local rule: given the other
edges, an edge is either forced to be absent or remains a $\Bernoulli(p)$ bit. The
positive inputs plant a uniformly random triangle $T$, and condition on it being the
unique one.

\subsection{Discussion}\label{sec:discussion}

\paragraph{Related work.} Our proof uses Zhandry's compressed oracle method~\cite{Zha19}, in the form adapted
to product distributions with biased bits~\cite{CMSZ21,HM23}, as the guide for the
calculation with independent bits, and Belovs's transfer operator
approach~\cite{Bel26} for the comparison of computations. Gilani and
Zhou~\cite{GZ26} used these tools to prove superlinear lower bounds for finding
certificates under product distributions, including the subgraph applications
mentioned above. The comparison is also related to the general adversary
method~\cite{HLS07}: the query error depends only on the matrix entries joining
inputs that differ at the queried coordinate. We estimate this error through
probability-weighted bounds on workspace vectors.

Recording arguments for input distributions conditioned to exclude
rare events were also developed by Gilani, Wang, Wu, and Zhou~\cite{GWWZ26} for
triangle finding in the edge list model. Here, by contrast, the conditioning event
of being triangle-free has exponentially small probability under the product distribution,
and our estimates analyze conditional resampling directly under the dependent
distribution. Sums of conditional variances also arise in probabilistic analyses of
conditional resampling; see, for example,~\cite[Section 1.1]{CMT15}.

\paragraph{What limits the exponent.}
The bound of \cref{thm:fixed-graph} improves as the density grows, and the density is
limited by our use of the conditioned distribution. The probability estimate
\eqref{eq:G-probability} and the interaction estimates of \cref{lem:triangle-star}
both need $\Delta p^2$ to be a small constant. This condition keeps the interaction
between conditional averages small enough to bound sums of conditional variances
as we bounded sums of record weights for independent bits. The calculation of
\cref{sec:overview} itself does not limit the density. At the largest allowed density,
\cref{thm:main} gives the lower bound $\Omega(\sqrt M\,\Delta^{-5/12})$.
Since $M\le N\Delta^2$ and $\Delta\le N$, we have
$\sqrt M\,\Delta^{-5/12}\le N^{13/12}$.
Thus, up to constant factors, the complete tripartite graph is the best choice.

\paragraph{Tightness for sparse graphs.}
Smaller densities give bounds for sparse graphs, which
are tight up to polylogarithmic factors in a range of edge counts.

For $G=K_{n,n,n}$, \Cref{thm:main} applies for every
$p$ with $C\log n/n\le p\le1/(10\sqrt{2n})$, for an absolute constant $C$, and gives
$t=\Omega(n^{7/6}p^{1/6})$. Let $N\coloneqq3n$ and $\bar m\coloneqq4n^2p$, and let
$\mathcal C$ be the class of graphs with at most $\bar m$ edges and maximum degree at
most $13\bar m/N$. Write $\mu_+$ for the positive input distribution. With the planted
triangle fixed, removing outside edges
cannot create another triangle. Thus \cref{lem:domination} bounds the upper tails
of the outside edges by those of independent Bernoulli bits. Accounting for the
three planted edges, which increase any degree by at most two, Chernoff bounds give
$\mu_+(\mathcal C)=1-o(1)$. An algorithm succeeding with probability at least
$2/3$ on $\mathcal C$, run with the same query limit on every input, therefore has
average success $\sigma\ge(2/3)\mu_+(\mathcal C)=2/3-o(1)$, while the exact
verifier rejects every triangle-free input. Since
$n^{7/6}p^{1/6}=\Theta(N^{5/6}\bar m^{1/6})$, finding a triangle in
graphs from $\mathcal C$ requires
\[
 \Omega\bigl(N^{5/6}\,\bar m^{1/6}\bigr)\ \text{queries}
 \qquad\text{for }N\log N\lesssim\bar m\lesssim N^{3/2},
\]
even when the triangle is unique.
Applying the extended learning graph of Carette, Lauri\`ere, and
Magniez~\cite[Theorem 6]{CLM20} gives triangle detection on $\mathcal C$ using
$O\bigl(N^{5/6}(\bar m\log N)^{1/6}+\bar m/\sqrt N\bigr)$ queries.{\interfootnotelinepenalty=10000\footnote{Add
 a known disjoint bipartite graph on two parts of size $N$, regular of degree
 $\lceil\bar m/N\rceil$. The padded graph has $3N$ vertices, between $\bar m$
 and $3\bar m$ edges, and maximum and root-mean-square degrees $O(\bar m/N)$.
 Choose the subset size in their learning graph to be
 $\lceil(3N)^{4/3}/(\bar m\log(3N))^{1/3}\rceil$. For sufficiently large $N$
 in the stated range, it lies between $(3N)^2/\bar m$ and $3N$, so Theorem~6
 applies to every padded input and gives the displayed bound uniformly.
 The added component contains no triangles and requires no input queries.}}
The second term is smaller than the first when $\bar m\le N^{8/5}$.
Finding reduces to detection at a polylogarithmic cost: repeatedly partition the
remaining vertices into four nearly equal parts and test which part can be removed while
retaining a triangle. Keeping removed vertices as isolated vertices preserves
both bounds defining $\mathcal C$. Therefore \cref{thm:main} determines the query
complexity of triangle finding on this class of graphs up to polylogarithmic factors.
In particular, at $\bar m\approx N^{3/2}$, the exponent $13/12$ is optimal
for this input distribution up to polylogarithmic factors.

\paragraph{Open problems.}
The main question is the gap between $N^{13/12}$ and the upper bound $N^{5/4}$. By the
previous paragraph, closing it requires negative inputs that are denser, or more
structured, than a triangle-free Bernoulli graph at density $\Theta(N^{-1/2})$. At higher
densities, the interaction terms in our conditional resampling estimates are no
longer guaranteed to be small. Extending the argument would require controlling
these stronger interactions. The estimates apply to any family of forbidden all-ones
patterns, and it would be interesting to find other problems where comparison with a
conditioned distribution is the natural route to a lower bound.

\paragraph{Organization.} \Cref{sec:overview} explains the proof through the
calculation with independent bits and lists the facts that must be reproved for the
triangle-free distribution. \Cref{sec:preliminaries} fixes notation. \Cref{sec:tools}
proves the comparison theorem and the estimates for conditional resampling.
\Cref{sec:triangle} proves the lower bound for triangle
finding. %

%% file: figures/bound-timeline.tex
\begingroup
\definecolor{timelineaccent}{RGB}{55,91,122}
\colorlet{timelineink}{black!68}
\begin{tikzpicture}[x=.43cm,y=9.3cm,font=\footnotesize,
  milestone/.style={circle,fill=timelineink,inner sep=1.3pt},
  bound/.style={timelineink,line width=1.15pt,line join=round},
  leader/.style={black!45,line width=.45pt},
  lab/.style={inner sep=3pt,text=black}]
\fill[black!5] (1,{1/2}) -| (8,{10/7-1}) -| (10,{13/10-1})
  -| (17,{35/27-1}) -| (18,{9/7-1}) -| (19,{1/4})
  -- (31,{1/4}) -- (31,0) -- (1,0) -- cycle;
\foreach \h in {{1/12},{1/4}}{
  \draw[black!18,densely dashed,line width=.35pt] (0,\h) -- (31,\h);
}
\draw[black!65,line width=.5pt,->,>=stealth] (0,-.035) -- (0,.56);
\node[lab,anchor=south west] at (-.2,.56) {Exponent of $N$};
\draw[black!65,line width=.5pt,->,>=stealth] (0,-.035) -- (32.2,-.035);
\node[lab,anchor=west] at (32.2,-.035) {Year};
\foreach \e/\elabel in {1/1,{13/12}/{13/12},{5/4}/{5/4},{3/2}/{3/2}}{
  \draw[black!65,line width=.5pt] (-.16,{\e-1}) -- (.16,{\e-1});
  \node[lab,anchor=east] at (-.2,{\e-1}) {$\elabel$};
}
\foreach \yr in {1995,2000,2005,2010,2015,2020,2026}{
  \pgfmathtruncatemacro{\yearx}{\yr-1995}
  \draw[black!65,line width=.5pt] (\yearx,-.042) -- (\yearx,-.028);
  \node[lab,anchor=north] at (\yearx,-.042) {\yr};
}
\draw[bound] (1,{1/2}) -| (8,{10/7-1}) -| (10,{13/10-1})
  -| (17,{35/27-1}) -| (18,{9/7-1}) -| (19,{1/4})
  -- (31,{1/4});
\foreach \x/\e in {1/{3/2},8/{10/7},10/{13/10},17/{35/27},18/{9/7},19/{5/4},22/{5/4},31/{5/4}}{
  \node[milestone] at (\x,{\e-1}) {};
}
\node[lab,anchor=south west] at (1,{1/2}) {\cite{Gro96}};
\node[lab,anchor=south west] at (8,{10/7-1}) {\cite{Sze03}};
\node[lab,anchor=south west] at (10,{13/10-1}) {\cite{MSS07}};
\draw[leader] (17,{35/27-1}) -- (17,.35) -- (15.7,.35);
\node[lab,anchor=south] at (15.7,.35) {\cite{Bel12}};
\draw[leader] (18,{9/7-1}) -- (18,.41) -- (21.2,.41);
\node[lab,anchor=south] at (21.2,.41) {\cite{LMS13,JKM13}};
\node[lab,anchor=north west] at (19,{1/4}) {\cite{LG14}};
\node[lab,anchor=south west] at (22,{1/4}) {\cite{CLM20}};
\node[lab,anchor=south east] at (31,{1/4}) {\cite{Kin26}};
\node[lab,anchor=north,font=\footnotesize\itshape] at (4.5,.49) {Upper bound};
\draw[bound] (1,0) -- (31,0);
\node[lab,anchor=south west] at (2,0)
  {Lower bound from unstructured search~\cite{BBBV97}};
\draw[timelineaccent,line width=1.6pt,->,>=stealth]
  (31,0) -- (31,{1/12});
\node[milestone,fill=timelineaccent,inner sep=1.7pt] at (31,{1/12}) {};
\node[lab,anchor=east,text=timelineaccent,font=\footnotesize\bfseries]
  at (30.5,{1/12}) {This work};
\end{tikzpicture}
\endgroup

%% file: sections/overview.tex
\section{Proof overview}\label{sec:overview}

Throughout, $G$ is a fixed graph known to the algorithm, and the input is a
subgraph of $G$ given by $\vx\in\{0,1\}^I$ with $I\coloneqq E(G)$. Write
$m\coloneqq|I|$, $\Delta$ for the maximum degree of $G$, $\st(w)\subseteq I$ for
the set of edges of $G$ at a vertex $w$, $\Triangles$ for the set of triangles of
$G$, each identified with its three edges, $V(T)$ for the vertex set of
$T\in\Triangles$, and $M\coloneqq|\Triangles|$. We call the elements of
$\Triangles$ \emph{candidates}. The edge density is $p$. For
the main result $G=K_{n,n,n}$, so that $m=3n^2$, $\Delta=2n$, $M=n^3$, and
$p=\Theta(n^{-1/2})$.

\subsection{The comparison method and the input distributions}\label{sec:overview-comparison}

For an input distribution $\rho$, the joint state of an algorithm after $t$ queries is
$\ket{\Psi_t}=\sum_{\vx}\sqrt{\rho(\vx)}\ket{\psi_t^{\vx}}\otimes\ket\vx$, where
$\ket{\psi_t^{\vx}}$ is its workspace state on input $\vx$. More generally, a
\emph{joint vector} is $\vf=\sum_{\vx}\sqrt{\rho(\vx)}\,\vf_{\vx}\otimes\ket\vx$
with arbitrary workspace vectors $\vf_{\vx}$, and
$\ket\rho=\sum_{\vx}\sqrt{\rho(\vx)}\ket\vx$ is the initial state of the
input register. A query to an edge $e$ multiplies the workspace vector at
$\vx$ by $(-1)^{x_e}$ in the corresponding branch of the workspace.
We denote the corresponding operator acting on the input register by $\Phase_e$.

Let $\rho_-$ be a distribution on negative inputs and $\rho_+$ a distribution on positive inputs, with
the planted triangle kept as a label that the algorithm cannot access. Following
Belovs's transfer viewpoint~\cite{Bel26}, we look for a linear map $\Ups$ from the
negative to the positive input register with two properties:
\begin{enumerate}[label=(\roman*)]
\item $\Ups\ket{\rho_-}=\ket{\rho_+}$, so the two computations agree before the
first query;
\item each query changes the comparison by at most $\varepsilon$ on the states that
actually occur:
\[
 \norm{(\Oracle^+\Ups-\Ups\Oracle^-)\ket{\Psi^-}}\le\varepsilon ,
\]
where $\Oracle^\pm$ apply the query to the joint states under the two distributions.
\end{enumerate}
Input-independent unitaries commute with $\Ups$, so the errors add up:
$\norm{\ket{\Psi^+_t}-\Ups\ket{\Psi^-_t}}\le t\varepsilon$. Append three queries
that verify the proposed triangle. The verifier rejects every negative input, and
its acceptance projector acts on the workspace, hence commutes with $\Ups$.
Therefore the average success probability satisfies $\sqrt\sigma\le(t+3)\varepsilon$
(\cref{lem:transfer}).

To bound $\varepsilon$, write the part of the negative state just before a query with phase
register $\ket{1}$ as $\sum_e\ket{e,1}\otimes\vf_e$. These \emph{query components} satisfy
$\sum_e\norm{\vf_e}^2\le1$, and each has infinity norm
$\norm{\vf_e}_\infty\coloneqq\max_{\vx}\norm{(\vf_e)_{\vx}}\le1$, because the workspace
state is a unit vector on every input. The query error is the orthogonal sum of the
vectors $(\Phase^+_e\Ups-\Ups\Phase^-_e)\vf_e$. Suppose that, for every query index
$e$ and every negative joint vector $\vf$,
\[
 \norm{(\Phase^+_e\Ups-\Ups\Phase^-_e)\vf}^2
 \le\alpha\norm\vf^2+\beta\norm\vf_\infty^2.
\]
Summing over $e$ bounds the squared query error by
$\alpha\sum_e\norm{\vf_e}^2+\beta\sum_e\norm{\vf_e}_\infty^2
\le\alpha+m\beta$.
The terms $\beta\norm{\vf_e}_\infty^2$ account for inputs of small total
probability. Their maxima can occur at different inputs for different $e$,
which is why their sum carries the factor $m$.

The simplest instance is the hybrid argument of Bennett, Bernstein, Brassard, and
Vazirani~\cite{BBBV97}. For search on $m$ bits, with negative input $0^m$ and
positive input $\boldsymbol\delta_i$ with a single one at a uniformly random
position $i$, the map $\Ups\ket{0^m}\coloneqq m^{-1/2}\sum_i\ket{\boldsymbol\delta_i}$ satisfies
$(\Phase^+_j\Ups-\Ups\Phase^-_j)\ket{0^m}=-2m^{-1/2}\ket{\boldsymbol\delta_j}$, so
$\varepsilon=2/\sqrt m$. We will construct $\Ups$ with $\varepsilon\approx n^{-13/12}$.

\paragraph{The input distributions.}\label{sec:overview-laws}
Our eventual error bound decreases with \(p\) after optimizing the averaging parameter, so we seek a triangle-free input distribution with \(p\) as large as our estimates permit. A dense graph with independent edges, however, has many
triangles: at $p=\Theta(n^{-1/2})$, a random subgraph of $K_{n,n,n}$ has about
$n^{3/2}$ of them, and Grover search over triples finds one with
$O(n^{3/4})$ queries. The verifier step above needs negative inputs with no triangle at all. We
therefore use
\[
 \nu\coloneqq\Bernoulli(p)^{\otimes I},\qquad
 \mu_-\coloneqq\nu(\,\cdot\mid\text{no triangle}),
\]
and, for $T\in\Triangles$, the distribution $\mu_T$ that sets the edges of $T$ to one and
samples the other edges from $\nu$ conditioned on $T$ being the unique triangle.
The positive distribution $\mu_+$ chooses $T$ uniformly. Conditioning makes the edges
dependent, but the dependence is local. Given all other edges, an edge $uv$ is
either forced to be absent, because some $w$ has $uw$ and $vw$ present, or it is a
$\Bernoulli(p)$ bit. Moreover, planting $T$ into a negative input creates another
triangle only through a present path of length two that closes an edge of $T$.
There are at most $3\Delta$ such paths, each present with probability at most $p^2$
by \cref{lem:domination},
so planting changes little as long as $\Delta p^2$ is a small constant.

\subsection{One candidate: records and the four-input map}\label{sec:overview-local}

Until \cref{sec:overview-dictionary}, we replace $\mu_-$ by $\nu$ and $\mu_T$ by the
distribution $\nu_T$ that sets the edges of $T$ to one and keeps the other edges independent,
which plants $T$ without conditioning. We first describe what queries do for
independent bits, and then construct the comparison for a single candidate $T$.

\paragraph{Records.}\label{sec:overview-records}
For independent bits, the following orthonormal basis of each
coordinate of the input register makes queries transparent:
\[
 \ket{\omega_p}\coloneqq\sqrt{1-p}\ket0+\sqrt p\ket1,
 \qquad
 \ket{\eta_p}\coloneqq\sqrt p\ket0-\sqrt{1-p}\ket1.
\]
The initial input state is $\ket\nu=\ket{\omega_p}^{\otimes I}$. For $S\subseteq I$,
let $\ket S$ be the product state with $\ket{\eta_p}$ on the coordinates in $S$ and
$\ket{\omega_p}$ elsewhere, and expand a joint vector as
$\vf=\sum_S\va_S\otimes\ket S$ with workspace vectors $\va_S$. In the language of
Zhandry's compressed oracle~\cite{Zha19}, adapted to biased bits~\cite{CMSZ21,HM23},
the term $\va_S\otimes\ket S$ has \emph{records} on $S$.

Two operators connect records with workspace vectors. Let $\Cond_i$ average
the workspace vector over bit $i$, so that
$(\Cond_i\vf)_{\vx}=(1-p)\vf_{\vx^{i\leftarrow0}}+p\vf_{\vx^{i\leftarrow1}}$, and put
$\Deriv_i\coloneqq\identity-\Cond_i$. On coordinate $i$, $\Cond_i$ is the projection
onto $\ket{\omega_p}$ and $\Deriv_i$ the projection onto $\ket{\eta_p}$. Thus
$\Deriv_i$ fixes $\ket S$ if $i\in S$ and annihilates it otherwise: it projects onto
the components with a record at $i$. Its squared norm has two readings:
\begin{equation}\label{eq:ov-variance}
 \norm{\Deriv_i\vf}^2=\sum_{S\ni i}\norm{\va_S}^2
 =\E_{\vX\sim\nu}\norm{\vf_{\vX}-(\Cond_i\vf)_{\vX}}^2.
\end{equation}
The first sum is the weight of the records at $i$. In the second expression,
$(\Cond_i\vf)_{\vX}$ is the conditional mean with all other coordinates fixed,
so the expectation is the conditional variance averaged over those coordinates.
This expression also applies to dependent bits and is the quantity used in the
actual proof.

A query to $e$ acts on coordinate $e$ alone, by the matrix
$\bigl(\begin{smallmatrix}1-2p&2\sqrt{p(1-p)}\\2\sqrt{p(1-p)}&2p-1\end{smallmatrix}\bigr)$
in the basis $(\ket{\omega_p},\ket{\eta_p})$. Its off-diagonal entries create or remove
a record at $e$, and they are captured by the commutator
\begin{equation}\label{eq:ov-creation}
 \Comm_e\coloneqq[\Deriv_e,\Phase_e]
 =2\sqrt{p(1-p)}\,\bigl(\ket{\eta_p}\bra{\omega_p}-\ket{\omega_p}\bra{\eta_p}\bigr)_e,
 \qquad\norm{\Comm_e}=2\sqrt{p(1-p)}.
\end{equation}
Since $\Phase_e$ commutes with $\Deriv_i$ for $i\ne e$ and
$\Phase_e\Deriv_e\Phase_e-\Deriv_e=\Phase_e\Comm_e$, each query increases the
expected number of records $\langle\sum_i\Deriv_i\rangle$ by at most $2\sqrt{p(1-p)}$.
Both facts also hold for the conditioned distribution
(\cref{sec:conditioned-law}), so this
\emph{record budget} does too.

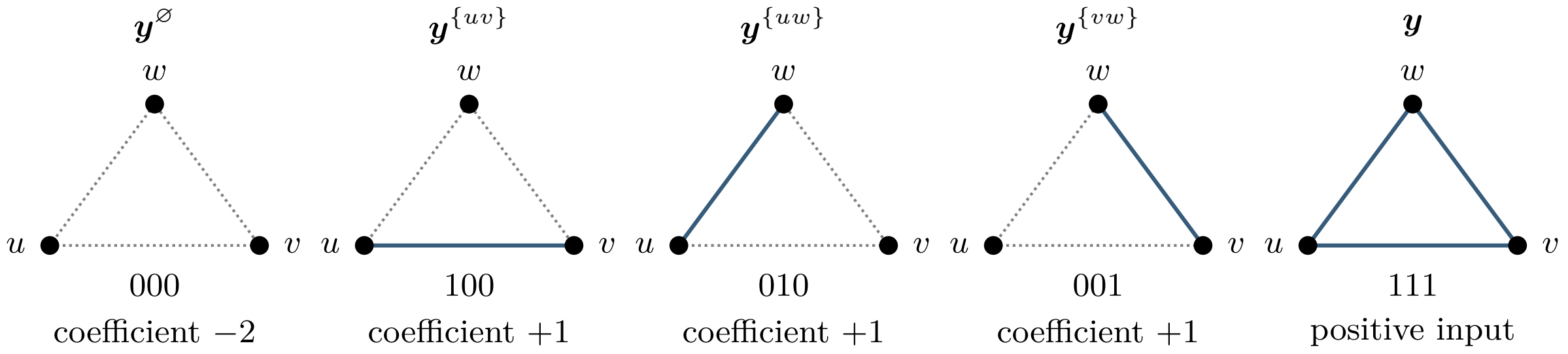
\begin{figure}[htbp]
 \centering
 \resizebox{\linewidth}{!}{\input{figures/four-inputs}}
 \caption{The four-input map for a positive input $\vy$ whose unique triangle is
 $T=\{uv,uw,vw\}$. The four inputs on the left are negative and agree with $\vy$
 outside $T$. Solid edges are present and dotted edges are absent; the bit order is
 $uv,uw,vw$. The coefficients $-2,+1,+1,+1$ are determined by
 $\Local_T\ket\nu=\ket{\nu_T}$ and
 $\Local_T\Phase_e^-\ket\nu=-\ket{\nu_T}$ for each $e\in T$.}
 \label{fig:four-inputs}
\end{figure}

\paragraph{The four-input map.}
The $T$-component of $\Ups$ will assign to each positive input containing $T$
a linear combination of workspace vectors at negative inputs. We need to choose
both the inputs and the coefficients in this combination. First consider a positive
input $\vy$ whose unique triangle is $T$. For $A\subseteq T$, let $\vy^A$ agree
with $\vy$ outside $T$ and contain exactly the edges of $A$ inside $T$. For
$A\ne T$ this input does not contain $T$, and under the distributions of
\cref{sec:overview-laws} it is negative. The map should satisfy
$\Ups\ket{\rho_-}=\ket{\rho_+}$ and give the same result whether an edge of $T$
is queried before or after applying the map, starting from the initial state.
Since each such edge is present in $\vy$, querying it after applying the map gives
phase $-1$. The linear combination of its query phases on the negative inputs must
therefore also equal $-1$. Each present edge of
$T$ makes an input rarer by a factor $p/(1-p)$, and the error bounds will pay for the
rarity of the inputs used. We therefore use the most probable negative assignments
$000,100,010,001$ of the three edges (\cref{fig:four-inputs}), with coefficients
$\lambda_0,\dots,\lambda_3$. The condition
$\Ups\ket{\rho_-}=\ket{\rho_+}$ requires $\sum_k\lambda_k=1$. A query
to the first edge of $T$ has phase $-1$ on $\vy$ and, among the four inputs, phase
$-1$ only on $100$. Matching the phase at $\vy$ therefore requires
$\lambda_0-\lambda_1+\lambda_2+\lambda_3=-1$, that is, $\lambda_1=1$. The other two
edges give $\lambda_2=\lambda_3=1$, and then $\lambda_0=-2$:
\begin{equation}\label{eq:ov-local}
 (\Local_T\vf)_{\vy}\coloneqq\sum_{i\in T}\vf_{\vy^{\{i\}}}-2\vf_{\vy^\varnothing},
\end{equation}
where the same rule is used for every $\vy\in\supp\nu_T$ in the
calculation for independent bits, and the joint vector $\Local_T\vf$ carries
the weights of $\nu_T$. For every $e\in T$, it satisfies
$\Local_T\Phase_e^-\ket\nu=-\ket{\nu_T}=\Phase_e^+\Local_T\ket\nu$.

In the computational basis on the three bits of $T$, $\Local_T$ is the
functional $(\bra{100}+\bra{010}+\bra{001})/\sqrt{p(1-p)^2}-2\bra{000}/\sqrt{(1-p)^3}$
with the fixed bits of the planted triangle suppressed. The denominators remove
the square root probability weights on $T$. The identities
$\bra0/\sqrt{1-p}=\bra{\omega_p}+\sqrt{p/(1-p)}\,\bra{\eta_p}$ and
$\bra1/\sqrt p=\bra{\omega_p}-\sqrt{(1-p)/p}\,\bra{\eta_p}$ give
\begin{equation}\label{eq:ov-local-records}
 \begin{gathered}
 \Local_T\ket S=c_{|S\cap T|}\ket{S\setminus T},\\
 c_0=1,\qquad c_1=-\sqrt{(1-p)/p},\qquad c_2=-2-\frac{p}{1-p},\qquad
 c_3=-3\sqrt{\frac{p}{1-p}}-2\Bigl(\frac{p}{1-p}\Bigr)^{3/2}.
 \end{gathered}
\end{equation}
Group the record components by their remaining record set $S\setminus T$. Each group
contains at most eight components, with coefficient $c_0=1$ for the component
having no record in $T$ and coefficients of order $O(p^{-1/2})$ for the others.
Different groups remain orthogonal after applying $\Local_T$.
Applying Cauchy--Schwarz within each group and summing gives
\begin{equation}\label{eq:ov-local-norm}
 \norm{\Local_T\vg}^2\le c\norm\vg^2+\frac cp\sum_{i\in T}\norm{\Deriv_i\vg}^2 .
\end{equation}
The factor $1/p$ is paid only by records in $T$. In particular
$\norm{\Local_T}^2\le c/p$.

\paragraph{The query error.}
For $e\notin T$ the four inputs agree with $\vy$ on $e$, so
$\Phase^+_e\Local_T=\Local_T\Phase^-_e$. For $e\in T$, the query gives phase $-1$
on $\vy$ and $\vy^{\{e\}}$, and phase $+1$ on the other three negative inputs.
Substituting these phases into \eqref{eq:ov-local} gives
\begin{equation}\label{eq:ov-local-error}
 \bigl((\Phase^+_e\Local_T-\Local_T\Phase^-_e)\vf\bigr)_{\vy}
 =-2\sum_{i\in T\setminus\{e\}}\bigl(\vf_{\vy^{\{i\}}}-\vf_{\vy^\varnothing}\bigr).
\end{equation}
The term $\vf_{\vy^{\{e\}}}$ has cancelled: whatever the query does at $e$, the error
depends only on the other two edges of $T$. In the record basis, the error vanishes
on components with no record on $T\setminus\{e\}$, multiplies a single record there by a
factor $O(p^{-1/2})$, and multiplies two records by $O(1)$. In either
language,
\begin{equation}\label{eq:ov-local-defect}
 \norm{(\Phase^+_e\Local_T-\Local_T\Phase^-_e)\vf}^2
 \le\frac cp\sum_{i\in T\setminus\{e\}}\norm{\Deriv_i\vf}^2 .
\end{equation}

\subsection{Many candidates: averaging over vertex stars}\label{sec:overview-smoothing}

\paragraph{Why averaging is needed.}\label{sec:overview-sum}
Let $\Ups_0\coloneqq M^{-1/2}\bigoplus_T\Local_T$. The factor $1/\sqrt M$
reflects choosing the planted triangle $T$ uniformly from $\Triangles$ in the
positive input distribution. Thus
\[
 \norm{\Ups_0\vf}^2=\frac1M\sum_{T\in\Triangles}\norm{\Local_T\vf}^2.
\]
Define $\Star_w\coloneqq\sum_{i\in\st(w)}\Deriv_i$. On a record state $\ket S$,
it counts the recorded edges incident to $w$:
$\Star_w\ket S=|S\cap\st(w)|\ket S$. Hence, for
$\vf=\sum_S\va_S\otimes\ket S$,
\[
 \ip\vf{\Star_w\vf}=\sum_S|S\cap\st(w)|\,\norm{\va_S}^2.
\]
For a normalized $\vf$, this is the expected number of recorded edges incident to
$w$. For $e=uv$, the candidates through $e$ are $\{uv,uw,vw\}$, one for
each common neighbour $w$, and each edge of $\st(u)\cup\st(v)$ is one of the
\emph{other} two edges for at most one of them. By \eqref{eq:ov-local-defect},
\begin{equation}\label{eq:ov-unsmoothed}
 \norm{(\Phase^+_e\Ups_0-\Ups_0\Phase^-_e)\vf}^2
 \le\frac c{pM}\,\ip\vf{(\Star_u+\Star_v)\vf}.
\end{equation}
The record budget bounds only the total number of records in expectation; it
does not prevent their concentration in the stars of $u$ and $v$. The bound
for a query to $uv$ can therefore depend on the entire record budget.
Since $\Star_u+\Star_v\preceq2\sum_{i\in I}\Deriv_i$, summing
\eqref{eq:ov-unsmoothed} over the query components of the state after $t$ queries
and applying the record budget bounds the squared error of the next query by
$\frac{O(t\sqrt p)}{pM}=O(t/(\sqrt p\,M))$.
The comparison argument adds the query errors in norm. Including the three
verification queries, for $t\ge1$ this gives
$\sqrt\sigma\lesssim t\sqrt{t/(\sqrt p\,M)}
\lesssim t^{3/2}p^{-1/4}M^{-1/2}$.
Hence we obtain only $t\gtrsim M^{1/3}p^{1/6}$, which is $n^{11/12}$ for $K_{n,n,n}$. On its own, the
four-input map thus gives only a sublinear lower bound for $K_{n,n,n}$.
The averaging introduced next will improve this bound by discounting components
with many records near the candidate.

\paragraph{The averaging.}
We precompose $\Local_T$ with an averaging operator $\Filter_T$ acting on the
negative input register. It should
\begin{enumerate}[label=(R\arabic*),leftmargin=3em]
\item satisfy $\Filter_T\ket\nu=\ket\nu$, so that
      $\Ups\ket{\rho_-}=\ket{\rho_+}$ still holds;
\item reduce the contribution of components with many records near $T$, while
allowing candidates that share a star to be bounded together using a single
averaging operator;
\item be an average over inputs, so that it does not increase the infinity norm,
which controls rare inputs in the actual proof;
\item have an explicit commutator with a query.
\end{enumerate}
Our choice is
\[
 \Energy_T\coloneqq\sum_{w\in V(T)}\Star_w=\sum_{i\in I}h_T(i)\Deriv_i,
 \qquad h_T(i)\coloneqq|i\cap V(T)|\in\{0,1,2\},
 \qquad \Filter_T\coloneqq(\identity+\Energy_T/R)^{-1},
\]
together with the averaging operators for individual stars
$\Filter_w\coloneqq(\identity+\Star_w/R)^{-1}$. In the record basis all of these
are diagonal. With $k_w(S)\coloneqq|S\cap\st(w)|$ and
$k_T(S)\coloneqq\sum_{w\in V(T)}k_w(S)$,
\begin{equation}\label{eq:ov-filter}
 \Filter_T\ket S=\frac R{R+k_T(S)}\ket S,
 \qquad
 \Filter_w\ket S=\frac R{R+k_w(S)}\ket S .
\end{equation}

The averaging operator $\Filter_T$ is defined by random-scan Gibbs sampling:
$(\Filter_T\vf)_{\vx}$ is the expected workspace vector at the final input of the walk
started at $\vx$. At each step, stop with probability $R/(R+\sum_ih_T(i))$;
otherwise choose an incidence between a vertex of $T$ and
an edge of its star uniformly, resample that edge from its conditional distribution, and
repeat. Resampling coordinate $i$ is the projection $\Cond_i$: it fixes $\ket S$ if
$i\notin S$ and annihilates it if $i\in S$. Thus $\Filter_T\ket S$ is $\ket S$ times
the probability that the walk stops before touching a recorded coordinate, which is
$R/(R+k_T(S))$: at each step, the probabilities of stopping and touching a
record are proportional to $R$ and $k_T(S)$, while updates to other coordinates
leave $\ket S$ unchanged. The sampling procedure gives (R1) and (R3).
The query commutator in (R4) is computed in
\eqref{eq:ov-filter-commutator}.
These three properties also hold under the conditioned distribution.

We average over all three stars so that the same map works for every query
index. Whichever edge $uv$ of $T$ is queried, its other two edges lie in $\st(u)$
and $\st(v)$, and these stars are part of the averaging for every candidate through
$uv$. Since $k_T(S)\ge k_u(S)$ and $k_T(S)\ge k_v(S)$,
\eqref{eq:ov-filter} lets us bound the contributions from all these candidates
using the averaging operators $\Filter_u$ and $\Filter_v$.

\paragraph{Summing the conditional variances.}
To prove (R2), we make this comparison precise and sum over each star using the
following three estimates:
\begin{align}
 &\text{commutation:} &&\Deriv_i\Filter_T=\Filter_T\Deriv_i;
 \label{eq:ov-S1}\\
 &\text{monotonicity:} &&\norm{\Filter_T\Deriv_i\vf}^2
   \le\ip{\Deriv_i\vf}{\Filter_w\Deriv_i\vf}\quad\text{for }w\in V(T);
 \label{eq:ov-S2}\\
 &\text{sum over a star:} &&\sum_{i\in\st(w)}\ip{\Deriv_i\vf}{\Filter_w\Deriv_i\vf}
   =\ip\vf{\Star_w\Filter_w\vf}\le R\norm\vf^2.
 \label{eq:ov-S3}
\end{align}
Equation~\eqref{eq:ov-S1} holds because both operators are diagonal in the record basis. Equation~\eqref{eq:ov-S2} holds
because $k_T\ge k_w$, so that $\Filter_T^2\preceq\Filter_T\preceq\Filter_w$. Equation~\eqref{eq:ov-S3}
holds because the $\Deriv_i$ commute, so the sum equals $\ip\vf{\Star_w\Filter_w\vf}$,
and $\Star_w\Filter_w=R(\identity-\Filter_w)$: a star with $k$ records, each damped by
$R/(R+k)$, costs $kR/(R+k)\le R$ in total.

For a queried edge $e=uv$, apply \eqref{eq:ov-S1} to each candidate $T\ni e$, then \eqref{eq:ov-S2} at $u$ for
the edge $uw$ and at $v$ for $vw$. As $T$ varies, its other edges $uw$ and $vw$ are
all distinct, so \eqref{eq:ov-S3} gives
\begin{equation}\label{eq:ov-through}
 \sum_{T\ni e}\ \sum_{i\in T\setminus\{e\}}\norm{\Deriv_i\Filter_T\vf}^2
 \le\sum_{i\in\st(u)}\ip{\Deriv_i\vf}{\Filter_u\Deriv_i\vf}
   +\sum_{i\in\st(v)}\ip{\Deriv_i\vf}{\Filter_v\Deriv_i\vf}
 \le2R\norm\vf^2 .
\end{equation}
With \eqref{eq:ov-local-defect}, the error of the four-input maps through $e$ is now
at most $cR/(pM)$ times $\norm\vf^2$, instead of up to $c\Delta/(pM)$.

\subsection{The price of averaging and the exponent}\label{sec:overview-price}

With averaging, the $T$-component of the query error is
\begin{equation}\label{eq:ov-split}
 \vg_T\coloneqq\Phase^+_e\Local_T\Filter_T\vf-\Local_T\Filter_T\Phase^-_e\vf
 =(\Phase^+_e\Local_T-\Local_T\Phase^-_e)\Filter_T\vf
   +\Local_T[\Phase^-_e,\Filter_T]\vf,
\end{equation}
and the squared error is $M^{-1}\sum_T\norm{\vg_T}^2$. The sum of squared norms of
the first terms is bounded using \eqref{eq:ov-local-defect} and
\eqref{eq:ov-through}. For the second term, the query sends $\ket S$ with $e\notin S$ to
$(1-2p)\ket S+2\sqrt{p(1-p)}\ket{S\cup\{e\}}$, so
\begin{equation}\label{eq:ov-filter-commutator}
 [\Phase_e,\Filter_T]\ket S=2\sqrt{p(1-p)}\left(\frac R{R+k}-\frac R{R+k+h}\right)
 \ket{S\cup\{e\}},
 \qquad k=k_T(S),\ h=h_T(e),
\end{equation}
and similarly with $S\setminus\{e\}$ when $e\in S$. The bracket is at most $h/R$.

\begin{figure}[htbp]
 \centering
 \resizebox{.9\linewidth}{!}{\input{figures/query-geometries}}
 \caption{A candidate triangle $T$ and a queried edge $uv$ of the fixed graph.}
 \label{fig:query-geometries}
\end{figure}
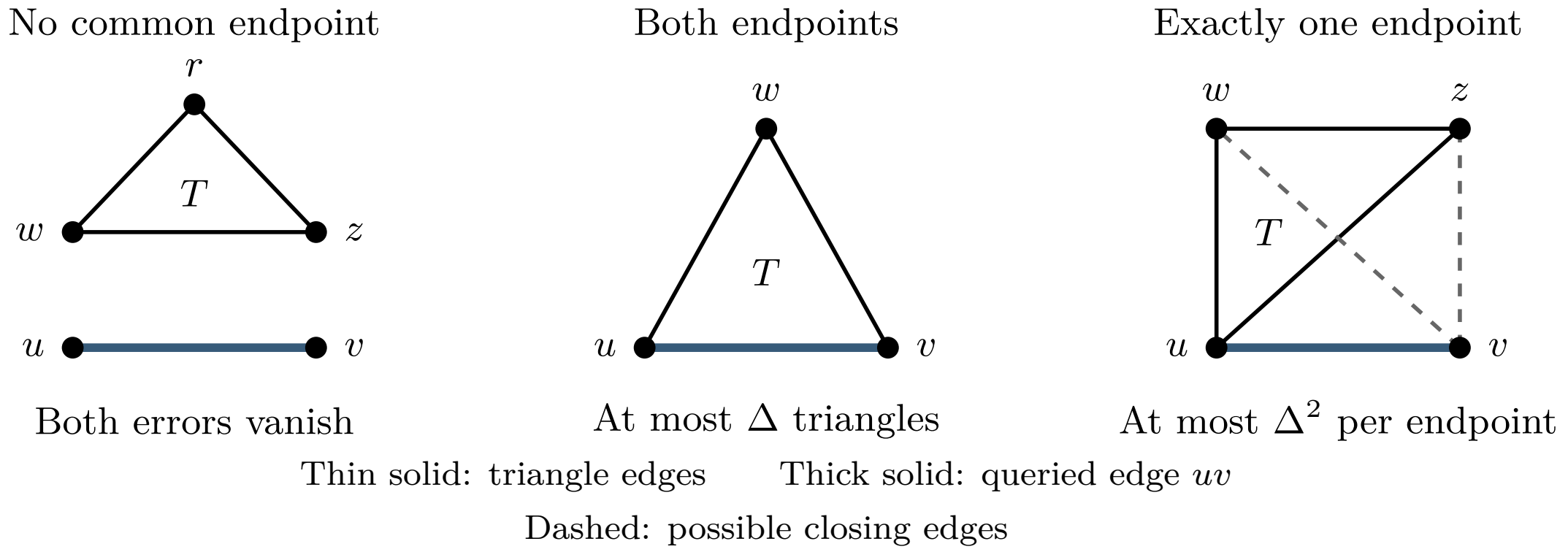

The commutator is nonzero exactly when $T$ contains an endpoint of $e$.
\Cref{fig:query-geometries} shows the possible positions. Write $\mathcal A_e$ for the
candidates containing $e=uv$, at most $\Delta$ of them, and $\mathcal B_{u,v}$ for those
containing $u$ but not $v$, at most $\binom\Delta2$ of them.

\paragraph{Both endpoints.} Here $h_T(e)=2$, so \eqref{eq:ov-filter-commutator}
gives $\norm{[\Phase_e,\Filter_T]}^2=O(p/R^2)$.
Together with $\norm{\Local_T}^2\le c/p$, this bounds the squared norm of the
second term in \eqref{eq:ov-split} by $O(\norm\vf^2/R^2)$ for each candidate.
Summing over $|\mathcal A_e|\le\Delta$ candidates and using
\eqref{eq:ov-local-defect} and \eqref{eq:ov-through} for the first term gives
\begin{equation}\label{eq:ov-A}
 \sum_{T\in\mathcal A_e}\norm{\vg_T}^2
 \le c\left(\frac Rp+\frac\Delta{R^2}\right)\norm\vf^2 .
\end{equation}

\paragraph{One endpoint.} Here $e\notin T$, so only the second term remains:
$\vg_T=\Local_T\boldsymbol q_T$ with $\boldsymbol q_T\coloneqq[\Phase_e,\Filter_T]\vf$ and $h_T(e)=1$. We
use \eqref{eq:ov-local-norm}. First,
$\sum_{T\in\mathcal B_{u,v}}\norm{\boldsymbol q_T}^2\le\binom\Delta2\frac{4p(1-p)}{R^2}\norm\vf^2
\le\frac{2\Delta^2p}{R^2}\norm\vf^2$.
Second, $[\Phase_e,\Filter_T]=R^{-1}\Filter_T\Comm_e\Filter_T$ since $h_T(e)=1$.
For $i\in T$ we have $i\ne e$, so by \eqref{eq:ov-S1} and $[\Deriv_i,\Comm_e]=0$,
$\Deriv_i\boldsymbol q_T=R^{-1}\Filter_T\Comm_e\Filter_T\Deriv_i\vf$. Hence the conditional
variances in \eqref{eq:ov-local-norm} contribute at most
$\frac{4c}{R^2}\sum_T\sum_{i\in T}\norm{\Filter_T\Deriv_i\vf}^2$.

To bound this sum, write each candidate as $T=\{uw,uz,wz\}$.
For the two edges $uw$ and $uz$, use the common star at $u$. Each such
edge occurs in at most $\Delta$ candidates, so \eqref{eq:ov-S2}--\eqref{eq:ov-S3} bound their total
contribution by $\Delta R\norm\vf^2$. For the opposite edge $wz$, use the
star of its first endpoint in a fixed vertex order. There are at most
$\Delta$ choices of this endpoint, since each is adjacent to $u$, and each
edge assigned to one of these stars belongs to at most one candidate
containing $u$. These edges contribute another $\Delta R\norm\vf^2$. Thus,
\begin{equation}\label{eq:ov-B-sum}
 \sum_{T\in\mathcal B_{u,v}}\ \sum_{i\in T}\norm{\Filter_T\Deriv_i\vf}^2
 \le2\Delta R\norm\vf^2 .
\end{equation}
Substituting these bounds into \eqref{eq:ov-local-norm} gives
\begin{equation}\label{eq:ov-B}
 \sum_{T\in\mathcal B_{u,v}}\norm{\vg_T}^2
 \le c\left(\frac{\Delta^2p}{R^2}+\frac\Delta R\right)\norm\vf^2 ,
\end{equation}
and the same holds for $\mathcal B_{v,u}$. Using only
$\norm{\Local_T}^2\le c/p$ would give $\Delta^2/R^2$ in place of
$\Delta^2p/R^2$ in \eqref{eq:ov-B}, losing the desired exponent.

\paragraph{Why use these averaging weights.} A sharp cutoff that keeps only components with
$k_T(S)\le R$ would also bound the error through $e$ by $cR/(pM)$. But its commutator
with a query has a term of size $2\sqrt{p(1-p)}$ on every component at the cutoff, and a
state with $R$ records in $\st(u)$ sits at the cutoff for all of the roughly
$\Delta^2$ candidates through $u$, costing order $\Delta^2p$ instead of $\Delta^2p/R^2$.

\paragraph{The exponent.}\label{sec:overview-exponent}
Adding \eqref{eq:ov-A} and two instances of \eqref{eq:ov-B} and dividing by $M$ gives
\begin{equation}\label{eq:ov-alpha}
 \frac1M\sum_{T\in\Triangles}\norm{\vg_T}^2\le
 \frac cM\left(\frac Rp+\frac\Delta R+\frac\Delta{R^2}+\frac{\Delta^2p}{R^2}\right)
 \norm\vf^2 .
\end{equation}
Taking $R=(\Delta p)^{2/3}\ge1$ balances $R/p$ and $\Delta^2p/R^2$;
the other terms are no larger since $\Delta p\ge1$.
For $K_{n,n,n}$ at $p=\Theta(n^{-1/2})$, $R=\Theta(n^{1/3})$: a
candidate is left essentially undamped until its stars carry about $n^{1/3}$ records,
while each star has about $\sqrt n$ present edges. The resulting error is
$\varepsilon\approx\Delta^{1/3}/(\sqrt M\,p^{1/6})$.
The bound on $\varepsilon$ decreases as $p$ grows. The density restriction comes
from the conditioning: planting locality and the
interaction estimates below need $\Delta p^2$ to be a small constant, so we take
$p=\Theta(\Delta^{-1/2})$. Then $\varepsilon\approx\Delta^{5/12}/\sqrt M$, which is $n^{-13/12}$ for
$K_{n,n,n}$, and $\sqrt\sigma\le(t+3)\varepsilon$ gives $t=\Omega(n^{13/12})$.

\subsection{From independent bits to the conditioned distribution}\label{sec:overview-dictionary}

The calculation above uses $\nu$, but the verifier step requires $\mu_-$.
We now take $\Cond_i$ to average coordinate $i$ conditionally on the other
coordinates under $\mu_-$,
and keep the same definitions of $\Deriv_i$, $\Filter_T$, and $\Filter_w$.
The projections $\Deriv_i$ need not commute, so there is no common product basis of
records. Two edge coordinates can interact when a third present edge completes
a triangle with them; the degrees of the sampled graph therefore bound the number
of interactions. The four-input map retains its formula for workspace vectors, but its
joint vectors now use the probability weights of $\mu_-$ and $\mu_T$.

\Cref{tab:dictionary} gives the replacements for the estimates under the product distribution.
The record budget still bounds the sum of conditional variances, and
\eqref{eq:ov-local-error} and the counting arguments remain valid.
The three steps that used commutation are replaced using the two-coordinate
estimate of \cref{lem:two-coordinate} and bounds on the sampled degrees.

\begin{table}[!htb]
 \centering
 \small
 \begin{tabular}{@{}>{\raggedright\arraybackslash}p{5.5cm}>{\raggedright\arraybackslash}p{\dimexpr\linewidth-8.85cm-4\tabcolsep\relax}>{\raggedright\arraybackslash}p{3.35cm}@{}}
  \toprule
  Fact for independent bits & Under the triangle-free distribution $\mu_-$ & Reference\\
  \midrule
  record weight $\norm{\Deriv_i\vf}^2$
   & conditional variance, averaged over the other coordinates
   & \cref{sec:averages}\\
  \midrule
  $\Filter_T\ket S=\frac R{R+k_T(S)}\ket S$
   & $\Filter_T$ averages workspace vectors by conditional resampling, satisfies
     $\Filter_T\ket{\mu_-}=\ket{\mu_-}$, and contracts
     $\norm\cdot_\infty$
   & \cref{lem:filter}\\
  \midrule
  \eqref{eq:ov-S2}
   & holds because inversion reverses order
   & \eqref{eq:triangle-average-order}\\
  \midrule
  $[\Phase_e,\Filter_T]=\frac{h_T(e)}R\Filter_T\Comm_e\Filter_T$
   & the operator identity holds verbatim, with $\norm{\Comm_e}\le2\sqrt p$
   & \eqref{eq:triangle-phase-average}\\
  \midrule
  norm bound \eqref{eq:ov-local-norm}; error bound \eqref{eq:ov-local-defect}
   & hold up to constants, by planting locality
   & \cref{lem:local-transfer}\\
  \midrule
  \eqref{eq:ov-S1}
   & $\norm{\Deriv_i\Filter_T\vf}^2\le4\norm{\Filter_T\Deriv_i\vf}^2
       +\frac pR\norm\vf^2+\text{rare}$
   & \eqref{eq:triangle-difference-average}\\
  \midrule
  \eqref{eq:ov-S3}
   & $\sum_{i\in\st(w)}\ip{\Deriv_i\vf}{\Filter_w\Deriv_i\vf}
       \le3R\norm\vf^2+\text{rare}$
   & \eqref{eq:triangle-star-energy}\\
  \midrule
  $[\Deriv_i,\Comm_e]=0$, $i\ne e$
   & $\norm{[\Deriv_i,\Comm_e]}\le2p$
   & \eqref{eq:triangle-query-difference-commutator}\\
  \midrule
  ---
   & terms $\beta\norm\vf_\infty^2$ from rare inputs contribute $m\beta$ to the squared query error
   & \cref{lem:transfer}; \eqref{eq:triangle-exceptional}\\
  \bottomrule
 \end{tabular}
 \caption{The estimates under the product distribution and their replacements under $\mu_-$, for
 $R\ge1$, $0<p\le1/(10\sqrt\Delta)$, and $\Delta p\ge1$.
 ``Rare'' denotes terms proportional to $\norm\vf_\infty^2$ and the bound on the degree tail
 in \eqref{eq:triangle-tail}; their coefficients are given in the cited estimates.}
 \label{tab:dictionary}
\end{table}

\Crefrange{sec:averages}{sec:coordinate-differences} establish the conditional resampling
estimates; \cref{sec:triangle} checks their hypotheses for the triangle-free distribution
and follows the calculation above. For candidates containing the queried edge,
replacing \eqref{eq:ov-S1} adds $O(\Delta/(MR))$ to the coefficient of $\norm\vf^2$.
For candidates containing just one endpoint, the additional contribution to this
coefficient is $O(\Delta^2p/(MR^2))$, using $R\ge1$.
Both orders already occur in \eqref{eq:ov-alpha}. The terms for rare inputs contain the
factor $e^{-c\Delta p}$. With the same choice of $R$, taking
$\Delta p\ge C_0\log(2m)$ for a sufficiently large
constant $C_0$ absorbs them even after summing over all $m$ query indices
by \eqref{eq:triangle-exceptional}. This gives the same exponent.

%% file: figures/four-inputs.tex
\begingroup%
\colorlet{ink}{black}%
\definecolor{accent}{RGB}{55,91,122}%
\colorlet{absent}{black!50}%
\begin{tikzpicture}[x=1cm,y=1cm,
  edge/.style={line width=1.1pt,accent},
  missing/.style={densely dotted,line width=.75pt,absent},
  vertex/.style={circle,fill=ink,inner sep=1.8pt},
  every node/.style={text=ink,font=\footnotesize}]
\foreach \j/\bits/\val/\coef in {
  0/000/{\boldsymbol y^{\varnothing}}/-2,
  1/100/{\boldsymbol y^{\{uv\}}}/+1,
  2/010/{\boldsymbol y^{\{uw\}}}/+1,
  3/001/{\boldsymbol y^{\{vw\}}}/+1,
  4/111/{\boldsymbol y}/{}}{
  \begin{scope}[xshift=\j*3cm]
    \coordinate (u) at (0,0);
    \coordinate (v) at (2,0);
    \coordinate (w) at (1,1.35);
    \draw[missing] (u)--(v)--(w)--cycle;
    \ifnum\j=1 \draw[edge] (u)--(v);\fi
    \ifnum\j=2 \draw[edge] (u)--(w);\fi
    \ifnum\j=3 \draw[edge] (v)--(w);\fi
    \ifnum\j=4 \draw[edge] (u)--(v)--(w)--cycle;\fi
    \node[vertex,label=left:$u$] at (u) {};
    \node[vertex,label=right:$v$] at (v) {};
    \node[vertex,label=above:$w$] at (w) {};
    \node at (1,2.10) {$\val$};
    \node at (1,-.38) {\bits};
    \ifnum\j<4 \node at (1,-.83) {coefficient $\coef$};\fi
    \ifnum\j=4 \node at (1,-.83) {positive input};\fi
  \end{scope}
}
\path[use as bounding box]
  ([xshift=-7pt,yshift=-7pt]current bounding box.south west)
  rectangle ([xshift=7pt,yshift=7pt]current bounding box.north east);
\end{tikzpicture}%
\endgroup

%% file: figures/query-geometries.tex
\begingroup%
\colorlet{ink}{black}%
\definecolor{accent}{RGB}{55,91,122}%
\colorlet{context}{black!60}%
\begin{tikzpicture}[x=1cm,y=1cm,
  edge/.style={line width=.95pt,ink},
  query/.style={line width=2pt,accent},
  closing/.style={dashed,line width=.95pt,context},
  vertex/.style={circle,fill=ink,inner sep=1.8pt},
  every node/.style={text=ink,font=\footnotesize}]
\begin{scope}
  \node at (1,2.65) {No common endpoint};
  \coordinate (u) at (0,0);
  \coordinate (v) at (2,0);
  \coordinate (w) at (0,.95);
  \coordinate (z) at (2,.95);
  \coordinate (r) at (1,2);
  \draw[query] (u)--(v);
  \draw[edge] (w)--(z)--(r)--cycle;
  \foreach \a/\pos in {u/left,v/right,w/left,z/right,r/above}
    \node[vertex,label=\pos:$\a$] at (\a) {};
  \node at (1,1.27) {$T$};
  \node at (1,-.6) {Both errors vanish};
\end{scope}
\begin{scope}[xshift=4.7cm]
  \node at (1,2.65) {Both endpoints};
  \coordinate (u) at (0,0);
  \coordinate (v) at (2,0);
  \coordinate (w) at (1,1.8);
  \draw[edge] (u)--(w)--(v);
  \draw[query] (u)--(v);
  \foreach \a/\pos in {u/left,v/right,w/above}
    \node[vertex,label=\pos:$\a$] at (\a) {};
  \node at (1,.62) {$T$};
  \node at (1,-.6) {At most $\Delta$ triangles};
\end{scope}
\begin{scope}[xshift=9.4cm]
  \node at (1,2.65) {Exactly one endpoint};
  \coordinate (u) at (0,0);
  \coordinate (v) at (2,0);
  \coordinate (w) at (0,1.8);
  \coordinate (z) at (2,1.8);
  \draw[edge] (u)--(w)--(z)--cycle;
  \draw[query] (u)--(v);
  \draw[closing] (v)--(w);
  \draw[closing] (v)--(z);
  \foreach \a/\pos in {u/left,v/right,w/above,z/above}
    \node[vertex,label=\pos:$\a$] at (\a) {};
  \node at (.43,.95) {$T$};
  \node at (1,-.6) {At most $\Delta^2$ per endpoint};
\end{scope}
\node[align=center,font=\scriptsize] at (5.7,-1.28)
  {Thin solid: triangle edges\qquad
   Thick solid: queried edge $uv$\\[3pt]
   Dashed: possible closing edges};
\path[use as bounding box]
  ([xshift=-7pt,yshift=-7pt]current bounding box.south west)
  rectangle ([xshift=7pt,yshift=7pt]current bounding box.north east);
\end{tikzpicture}%
\endgroup

%% file: sections/preliminaries.tex
\section{Preliminaries}
\label{sec:preliminaries}

\subsection{Notation}

Let $I$ be a finite set of input coordinates. An input is a bit string
$\vx\in\{0,1\}^I$. For $S\subseteq I$, write
$\vx_S\coloneqq(x_e)_{e\in S}$; the strings $\vzero_S$ and $\vone_S$
assign zero and one, respectively, to every coordinate of $S$.
We order bit strings coordinatewise. An event is increasing if changing
zeros to ones preserves membership in the event.

For a probability measure $\rho$ on a finite set $\Omega$, let
$\supp\rho\coloneqq\{\vx\in\Omega:\rho(\vx)>0\}$ and write $\E_\rho$
for expectation. We use $\rho(\,\cdot\mid A)$ only when $\rho(A)>0$,
and write $\one_A$ for the indicator of $A$. The distribution $\Bernoulli(p)$
assigns probability $p$ to one and $1-p$ to zero.

All Hilbert spaces are finite dimensional and complex, with inner products
conjugate-linear in the first argument. Script letters denote Hilbert
spaces and calligraphic letters denote sets or families. We write linear
operators as $\Op{A},\Op{B},\ldots$. We use kets for normalized pure
states and bold symbols for general vectors. Thus $\ket\psi$ has unit
norm, while $\vf$ and an operator image such as $\Op{A}\ket\psi$ need
not. The adjoint is denoted by $\dagger$, and
$\bra\psi\coloneqq\ket\psi^\dagger$.

Unmarked vector norms and inner products are Hilbert space norms and inner
products. For a linear map $\Op{A}$, its operator norm is
\[
 \norm{\Op{A}}\coloneqq
 \sup_{\vu\ne\vzero}\frac{\norm{\Op{A}\vu}}{\norm\vu}.
\]
We write $\identity$ for the identity operator and $\Pi$ for an orthogonal
projector, so $\Pi^\dagger=\Pi^2=\Pi$. For operators on the same space,
$[\Op{A},\Op{B}]\coloneqq\Op{A}\Op{B}-\Op{B}\Op{A}$.
For self-adjoint operators, $\Op{A}\preceq\Op{B}$ means that
$\Op{B}-\Op{A}$ is positive semidefinite, or equivalently that
$\ip\vu{(\Op{B}-\Op{A})\vu}\ge0$ for every $\vu$; the notation
$0\prec\Op{A}$ means that $\Op{A}$ is positive definite.

Unless stated otherwise, $\log$ denotes the natural logarithm.
Constants in asymptotic notation may depend on parameters explicitly held
fixed in the statement. The notation $\widetilde O$ suppresses polylogarithmic
factors. We use $c$ for positive constants that may change between occurrences,
specifying their allowed dependence, and $C$ for absolute constants.
Subscripted constants such as $C_0,C_1$ are fixed within the statement and
proof that use them.

\subsection{Queries, joint states, and workspace vectors}
\label{sec:joint-states}

The workspace $\W$ consists of an index register, a one-qubit phase register,
and the remaining registers $\Wrest$. Its basis states are written
$\ket{e,b}\otimes\ket w$, where $e\in I$, $b\in\{0,1\}$, and $w$
labels a basis of $\Wrest$. On input $\vx$, a query acts as
\begin{equation}\label{eq:oracle}
 \Oracle_{\vx}\bigl(\ket{e,b}\otimes\ket w\bigr)
 \coloneqq(-1)^{b x_e}\ket{e,b}\otimes\ket w.
\end{equation}
This is the usual bit oracle conjugated by a Hadamard on the phase
register, so it has the same query cost. All input-independent operations
are free. The branch $b=0$ allows us to pad a computation to a fixed
number of queries.

We purify private randomness and retain intermediate measurement outcomes
coherently. Starting from an input-independent pure state, an algorithm
alternates queries with input-independent unitaries and ends with a
measurement. Its workspace includes the output
register and any verification ancillas from the outset. On input $\vx$,
let $\ket{\psi_t^{\vx}}\in\W$ be the normalized state after $t$ queries.

For an input distribution $\rho$ on $\{0,1\}^I$, introduce the input register and
its initial state
\[
 \Input_\rho\coloneqq
     \operatorname{span}\{\ket\vx:\vx\in\supp\rho\},
 \qquad
 \ket\rho\coloneqq
     \sum_{\vx\in\supp\rho}\sqrt{\rho(\vx)}\ket\vx.
\]
The basis is orthonormal. We place the workspace before the input register,
and represent the computation by
\[
 \ket{\Psi_{t,\rho}}\coloneqq
 \sum_{\vx\in\supp\rho}
     \sqrt{\rho(\vx)}\ket{\psi_t^{\vx}}\otimes\ket\vx.
\]
This joint state is normalized because every $\ket{\psi_t^{\vx}}$ is
normalized. Tracing out the input register gives the algorithm's workspace
state averaged over $\rho$. The joint oracle applies $\Oracle_{\vx}$ to
the workspace on the component labelled by $\vx$. The input register is
used only for analysis and is accessible to the algorithm only through
these queries.

A search algorithm has bounded error on a promised class of positive
inputs if it returns a valid solution with probability at least $2/3$
on every input in the class. Its output on an input with no solution
is unrestricted. We will append an exact verifier when comparing these
two cases.

\paragraph{Workspace vectors and local operations.}
\label{sec:conditional-notation}
The proof also uses vectors obtained by applying linear maps to joint
states. For a fixed distribution $\rho$, every $\vf\in\W\otimes\Input_\rho$
has a unique expansion
\begin{equation}\label{eq:joint-vector}
 \vf=\sum_{\vx\in\supp\rho}
      \sqrt{\rho(\vx)}\,\vf_{\vx}\otimes\ket\vx,
 \qquad \vf_{\vx}\in\W.
\end{equation}
We call $\vf_{\vx}$ the \emph{workspace vector} associated with input $\vx$.
It does not include the factor $\sqrt{\rho(\vx)}$. Orthogonality gives
\[
 \ip\vf\vg=\E_{\vx\sim\rho}\ip{\vf_{\vx}}{\vg_{\vx}}_\W,
 \qquad
 \norm\vf^2=\E_{\vx\sim\rho}\norm{\vf_{\vx}}_\W^2.
\]
We call the maximum of their norms the \emph{infinity norm}:
\[
 \norm\vf_\infty\coloneqq
      \max_{\vx\in\supp\rho}\norm{\vf_{\vx}}_\W.
\]
This norm depends on the reference distribution $\rho$. For any event $A$, it bounds
the contribution of inputs in $A$ by
\[
 \sum_{\vx\in A\cap\supp\rho}\rho(\vx)\norm{\vf_{\vx}}_\W^2
 \le\rho(A)\norm\vf_\infty^2.
\]
For an actual computation, the workspace vectors are the unit
vectors $\ket{\psi_t^{\vx}}$, and hence
\[
 \norm{\ket{\psi_t^{\vx}}}=1\quad\text{for every }\vx,
 \qquad
 \norm{\ket{\Psi_{t,\rho}}}
   =\norm{\ket{\Psi_{t,\rho}}}_\infty=1.
\]
In particular, every query component has workspace vectors of norm at
most one, since it is obtained by projecting each $\ket{\psi_t^{\vx}}$.

We specify local operations by their action on the workspace vectors. The
resulting joint vector retains the same square root probability weights
unless a different distribution is specified for that vector. Operators acting on the input
register act as the identity on the workspace and therefore commute with
operations on the workspace. All estimates are uniform in the dimension
of $\W$.

In particular, define the phase operator $\Phase_e$ acting on the input
register by
\begin{equation}\label{eq:input-phase}
 (\Phase_e\vf)_{\vx}\coloneqq(-1)^{x_e}\vf_{\vx}.
\end{equation}
A controlled query applies $\Phase_e$ on components with query index $e$
and phase register $\ket{1}$, and acts as the identity when the phase
register is $\ket{0}$. Superscripts
$-$ and $+$ distinguish phase operators for two input distributions. A distribution may
also include a hidden label $a$, in which case we use basis states
$\ket{a,\vx}$ and the same square root probability convention. Queries
retain the label and read only $\vx$.

Finally, for $\vX\sim\rho$ on $\{0,1\}^I$, define conditional averaging
over coordinate $e$ by
\begin{equation}\label{eq:conditional-average}
 (\Cond_{\rho,e}\vf)_{\vx}
   \coloneqq\E_\rho\bigl[\vf_{\vX}\mid
       \vX_{I\setminus\{e\}}=\vx_{I\setminus\{e\}}\bigr],
 \qquad \vx\in\supp\rho.
\end{equation}
This operation replaces each workspace vector by its conditional
mean, with all coordinates except $e$ fixed. Write
\[
 \Deriv_{\rho,e}\coloneqq\identity-\Cond_{\rho,e}
\]
for the corresponding conditional difference. We omit $\rho$ from the
subscripts when the distribution is fixed. In \cref{sec:averages}, we write these
operators explicitly on the two possible values of the bit and prove
their projection properties.

%% file: sections/tools.tex
\section{Comparison and conditional resampling}\label{sec:tools}

We compare computations under two input distributions using a linear map in the
sense of Belovs~\cite{Bel26}. The map matches the initial states, and
bounding its error under each query lets us bound the algorithm's success
probability.

We use conditional resampling to average the workspace vectors before
applying the comparison map. The averaging reduces a weighted sum of
conditional variances while almost commuting with each query.
We bound these effects first for
arbitrary Boolean input distributions, then for Bernoulli bits conditioned to
avoid all-ones patterns. For the latter, the main step is to bound the
interaction between two conditional averages by the corresponding
conditional variances. The resulting estimates control sums of averaged
conditional variances and the error in moving a conditional difference
through an averaging operator.

\input{sections/comparison}
\input{sections/resampling}

%% file: sections/comparison.tex
\subsection{Comparing two computations}\label{sec:comparison}

Let $\rho_-$ and $\rho_+$ be two input distributions for a search problem on the
same Boolean coordinate set $I$. The negative distribution is supported on inputs
with no valid output. The positive distribution may include a hidden label, in
which case its input register has basis states $\ket{a,\vx}$ and the
oracle reads only $\vx$. The algorithm has no direct access to either
input register. Write $\X,\Y$ for the two input spaces and
$\ket{\rho_-},\ket{\rho_+}$ for their normalized probability
superpositions.

Recall the joint vector convention in \eqref{eq:joint-vector}. For a
workspace $\W$, a negative joint vector $\vf\in\W\otimes\X$ has
workspace vectors $\vf_{\vx}\in\W$, with
\[
 \vf=\sum_{\vx\in\supp\rho_-}
       \sqrt{\rho_-(\vx)}\,\vf_{\vx}\otimes\ket\vx.
\]
Its Hilbert norm and infinity norm are, respectively,
\[
 \norm\vf^2=\E_{\vx\sim\rho_-}\norm{\vf_{\vx}}^2,
 \qquad
 \norm\vf_\infty=
       \max_{\vx\in\supp\rho_-}\norm{\vf_{\vx}}.
\]

We seek a linear map $\Ups:\X\to\Y$ with
$\Ups\ket{\rho_-}=\ket{\rho_+}$. This equality makes the two
computations agree before any queries. The map acts on the input register
and as the identity on the workspace. Under either distribution, the phase operator
$\Phase_e^\pm$ acts on the corresponding input register, multiplying a
basis vector labelled by $\vx$ by $(-1)^{x_e}$.
The corresponding controlled oracle $\Oracle^\pm$ applies
$\Phase_e^\pm$ on components with query index $e$ and phase register
$\ket{1}$, and acts as the identity when the phase register is $\ket{0}$.

\begin{samepage}
\begin{theorem}[Comparison of computations]\label[theorem]{lem:transfer}
Suppose $\Ups\ket{\rho_-}=\ket{\rho_+}$ and there are nonnegative
numbers $\alpha_e,\beta_e$ such that, for every finite workspace $\W$
and every $\vf\in\W\otimes\X$,
\begin{equation}\label{eq:abstract-defect}
 \norm{(\Phase_e^+\Ups-\Ups\Phase_e^-)\vf}^2
 \le\alpha_e\norm\vf^2+\beta_e\norm\vf_\infty^2.
\end{equation}
Suppose an output can be verified exactly with $v$ queries, with
verification always rejecting on
negative inputs. Every $t$-query algorithm with average success probability
$\sigma$ under $\rho_+$ then satisfies
\begin{equation}\label{eq:comparison-conclusion}
 \sqrt\sigma\le(t+v)
       \left(\max_{e\in I}\alpha_e+\sum_{e\in I}\beta_e\right)^{1/2}.
\end{equation}
\end{theorem}
\end{samepage}

\begin{proof}
Let $\ket{\Psi^-}$ be an actual negative joint state just before a query,
with normalized workspace state $\ket{\psi^{\vx}}$ on each input.
Extract its components with query index $e$ and phase register $\ket{1}$:
\[
 \vf_e\coloneqq(\bra{e,1}\otimes\identity)\ket{\Psi^-},
 \qquad
 (\vf_e)_{\vx}=(\bra{e,1}\otimes\identity)\ket{\psi^{\vx}}.
\]
These vectors need not be normalized. Orthogonality of the index
register and normalization on each input give
\[
 \sum_e\norm{\vf_e}^2\le1,\qquad \norm{\vf_e}_\infty\le1.
\]
Components with phase register $\ket{0}$ contribute no error. The
comparison preserves the index and phase registers, so its errors for
distinct query indices are orthogonal. Thus \eqref{eq:abstract-defect} implies
\begin{equation}\label{eq:coherent-defect}
 \norm{(\Oracle^+\Ups-\Ups\Oracle^-)\ket{\Psi^-}}^2
 \le\varepsilon^2,
 \qquad
 \varepsilon^2\coloneqq\max_e\alpha_e+\sum_e\beta_e.
\end{equation}
Run the same circuit under the two distributions. Initially,
$\ket{\Psi_0^+}=\Ups\ket{\Psi_0^-}$ because the workspace is input
independent. If $\Unitary_j$ is the next input-independent unitary, then
\begin{align*}
 \ket{\Psi_{j+1}^+}-\Ups\ket{\Psi_{j+1}^-}
 ={}&\Unitary_j\Oracle^+\ket{\Psi_j^+}
       -\Ups\Unitary_j\Oracle^-\ket{\Psi_j^-}\\
 ={}&\Unitary_j\Oracle^+
       \bigl(\ket{\Psi_j^+}-\Ups\ket{\Psi_j^-}\bigr)
       +\Unitary_j(\Oracle^+\Ups-\Ups\Oracle^-)\ket{\Psi_j^-}.
\end{align*}
In the last equality, we commuted $\Ups$ with $\Unitary_j$ and added
and subtracted $\Unitary_j\Oracle^+\Ups\ket{\Psi_j^-}$.
The first term propagates the previous error by a unitary, and the second
has norm at most $\varepsilon$ by \eqref{eq:coherent-defect}. Induction gives
$\norm{\ket{\Psi_j^+}-\Ups\ket{\Psi_j^-}}\le j\varepsilon$.

Append the exact verifier coherently, padding to $v$ queries if necessary.
Let $\Piacc$ be its acceptance projector. Verification preserves the
original success probability and rejects every negative input, so
$\Piacc\ket{\Psi_{t+v}^-}=\vzero$. Since the projector acts only on
workspace, it commutes with $\Ups$. Consequently,
\[
 \sqrt\sigma=\norm{\Piacc\ket{\Psi_{t+v}^+}}
 =\norm{\Piacc\bigl(\ket{\Psi_{t+v}^+}
                         -\Ups\ket{\Psi_{t+v}^-}\bigr)}
 \le\norm{\ket{\Psi_{t+v}^+}-\Ups\ket{\Psi_{t+v}^-}}
 \le(t+v)\varepsilon.\qedhere
\]
\end{proof}

The two terms in \eqref{eq:abstract-defect} behave differently when the
query index is in superposition. For an actual state,
$\sum_e\norm{\vf_e}^2\le1$, so the contributions
$\alpha_e\norm{\vf_e}^2$ sum to at most $\max_e\alpha_e$.
We also have $\norm{\vf_e}_\infty\le1$ for every $e$, but these maxima
may occur on different inputs. Different inputs can place their entire
amplitude on different query indices, making
$\sum_e\norm{\vf_e}_\infty^2$ as large as $|I|$.
Thus the contributions $\beta_e\norm{\vf_e}_\infty^2$ are bounded
by $\sum_e\beta_e$.

In our application, $|I|=m$ and each $\beta_e$ contains a factor
$e^{-\Omega(\Delta p)}$. The hypothesis
$\Delta p\ge C_0\log(2m)$ makes this factor small enough to absorb
the additional factor $m$ from summing over query indices.

%% file: sections/resampling.tex
\subsection{Conditional variances and averaging}\label{sec:averages}

Let $\rho$ be a probability measure on $\{0,1\}^I$. All operators in this
subsection use $\rho$. Fix an assignment to the coordinates outside $i$ that has positive
probability. If both values of $x_i$ are allowed, let $q$ be the
conditional probability of $x_i=1$. Suppressing the fixed coordinates,
define
\[
 \ket{\omega_q}\coloneqq\sqrt{1-q}\ket0+\sqrt q\ket1,
 \qquad
 \ket{\eta_q}\coloneqq\sqrt q\ket0-\sqrt{1-q}\ket1.
\]
The restrictions of $\Cond_i$ and $\Deriv_i$ to these two inputs are
\[
 \Cond_i=\ket{\omega_q}\bra{\omega_q},
 \qquad
 \Deriv_i=\ket{\eta_q}\bra{\eta_q}.
\]
Write $\vf_0,\vf_1$ for the workspace vectors corresponding to
$x_i=0,1$, respectively, with the same fixed assignment outside $i$.
After suppressing the fixed outside coordinates and their common
square root probability weight, the joint vector is
\[
 \sqrt{1-q}\,\vf_0\otimes\ket0+\sqrt q\,\vf_1\otimes\ket1.
\]
The projectors act on the input bit of this vector. The operator $\Cond_i$
forgets the dependence of the workspace vectors on $x_i$, with the
outside coordinates fixed, while $\Deriv_i$ retains that dependence:
\begin{equation}\label{eq:conditional-coordinates}
 \begin{aligned}
 (\Cond_i\vf)_0=(\Cond_i\vf)_1
     &=(1-q)\vf_0+q\vf_1,\\
 (\Deriv_i\vf)_0&=q(\vf_0-\vf_1),\qquad
 (\Deriv_i\vf)_1=(1-q)(\vf_1-\vf_0).
 \end{aligned}
\end{equation}
Since $\ket{\omega_q}$ and $\ket{\eta_q}$ are orthonormal, these
restrictions are complementary orthogonal projections. If only one
value of $x_i$ is allowed, $\Cond_i$ acts as the identity and
$\Deriv_i$ is zero on that input. Hence
\[
 \Cond_i^\dagger=\Cond_i^2=\Cond_i,\qquad
 \Deriv_i^\dagger=\Deriv_i^2=\Deriv_i,\qquad
 \ip\vf{\Deriv_i\vf}=\norm{\Deriv_i\vf}^2.
\]
The squared norm $\norm{\Deriv_i\vf}^2$ is the conditional variance of
the workspace vectors, averaged over the outside coordinates:
\[
 \norm{\Deriv_i\vf}^2
 =\E_{\vX\sim\rho}
       \norm{\vf_{\vX}-(\Cond_i\vf)_{\vX}}^2
 =\E_{\vX\sim\rho}
       \norm{\vf_{\vX}
       -\E_\rho[\vf_{\vX}\mid\vX_{I\setminus\{i\}}]}^2.
\]
For nonnegative weights $(w_i)_{i\in I}$, put
\begin{equation}\label{eq:weighted-energy}
 w\coloneqq\sum_{i\in I}w_i,\qquad
 \Energy\coloneqq\sum_{i\in I}w_i\Deriv_i,
 \qquad
 \ip\vf{\Energy\vf}=\sum_{i\in I}w_i\norm{\Deriv_i\vf}^2.
\end{equation}
Thus $\ip\vf{\Energy\vf}$ is the weighted sum of conditional variances.
For the product distribution $\rho=\Bernoulli(p)^{\otimes I}$, we have $q=p$
for every outside assignment. In the basis of \cref{sec:overview-records},
$\Cond_i$ discards the components with a record at $i$, while $\Deriv_i$
selects them. Thus $\Energy$ counts records with weights $w_i$. For dependent bits, $q$
can depend on the outside assignment, and the projections $\Deriv_i$
need not commute.

Instead of resampling in a fixed order, we use conditional resampling to distribute
the updates randomly across the coordinates, reducing the weighted sum of
conditional variances while keeping the error introduced at each query index small.
Fix $R>0$.
Starting at $\vx\in\supp\rho$, stop with probability
$R/(R+w)$. Otherwise choose $i$ with probability $w_i/w$, resample it
from its conditional distribution under $\rho$, and repeat. When $w=0$, stop
immediately. Let $K$ count the resampling steps, with expectation
$w/R$, and let $\vX_K$ be the final input. Define the averaging operator $\Filter$ by
\[
 (\Filter\vf)_{\vx}\coloneqq
     \E[\vf_{\vX_K}\mid\vX_0=\vx].
\]
The joint vector $\Filter\vf$ retains the weights $\sqrt{\rho(\vx)}$.
Resampling from the conditional distribution keeps every input in the support of
$\rho$. The averaging operator satisfies $\Filter\ket\rho=\ket\rho$ and
contracts the infinity norm. The stopping rule also gives a useful formula for
comparing different choices of resampling coordinates. For $w>0$, one
step has averaging operator $\Avg\coloneqq w^{-1}\sum_iw_i\Cond_i$.
The averaging operator $\Filter$ is an average of powers of $\Avg$, so the two
operators commute. Conditioning on whether the procedure stops before
its first update gives
\[
 \Filter=\frac R{R+w}\identity+\frac w{R+w}\Filter\Avg.
\]
Since $\Energy=w(\identity-\Avg)$, we obtain
\begin{equation}\label{eq:filters}
 \Filter=(\identity+\Energy/R)^{-1}.
\end{equation}
The inverse exists because $\Energy$ is positive semidefinite.
Equation~\eqref{eq:filters} also holds when $w=0$.

\begin{lemma}[Properties of conditional resampling]\label[lemma]{lem:filter}
The averaging operator $\Filter$ is self-adjoint and satisfies
\begin{equation}\label{eq:filter-properties}
 0\preceq\Filter^2\preceq\Filter\preceq\identity,
 \qquad \Filter\ket\rho=\ket\rho,
 \qquad \norm{\Filter\vf}_\infty\le\norm\vf_\infty.
\end{equation}
If $\Energy'\succeq\Energy$ and
$\Filter'\coloneqq(\identity+\Energy'/R)^{-1}$, then
\begin{equation}\label{eq:filter-order}
 0\preceq(\Filter')^2\preceq\Filter'\preceq\Filter.
\end{equation}
For every operator $\Op{B}$ acting on the input register,
\begin{equation}\label{eq:resolvent-comm}
 [\Op{B},\Filter]=R^{-1}\Filter[\Energy,\Op{B}]\Filter.
\end{equation}
Moreover,
\begin{equation}\label{eq:averaged-energy}
 \ip{\Filter\vf}{\Energy\Filter\vf}\le\frac R4\norm\vf^2.
\end{equation}
All infinity norms in this lemma use $\rho$.
\end{lemma}

\begin{proof}
Since $\Energy$ is positive semidefinite, it has an orthonormal
eigenbasis with nonnegative eigenvalues. By \eqref{eq:filters},
$\Filter$ is diagonal in the same basis, with eigenvalues
$(1+\lambda/R)^{-1}\in(0,1]$ for $\lambda\ge0$.
Thus $\Filter$ is self-adjoint and
$0\preceq\Filter^2\preceq\Filter\preceq\identity$.

By \eqref{eq:conditional-average}, we have $\Cond_i\ket\rho=\ket\rho$
for every $i$. Thus $\Energy\ket\rho=0$, and \eqref{eq:filters} gives
$\Filter\ket\rho=\ket\rho$. Moreover, the triangle inequality gives,
for every $\vx\in\supp\rho$,
\[
 \norm{(\Filter\vf)_{\vx}}
 =\norm{\E[\vf_{\vX_K}\mid\vX_0=\vx]}
 \le\E[\norm{\vf_{\vX_K}}\mid\vX_0=\vx]
 \le\norm\vf_\infty.
\]
Taking the maximum over $\vx$ proves
$\norm{\Filter\vf}_\infty\le\norm\vf_\infty$.

Inversion reverses the order of positive definite operators. Applying this to
$\identity+\Energy/R\preceq\identity+\Energy'/R$ proves
\eqref{eq:filter-order}.

For \eqref{eq:resolvent-comm}, use $\Filter^{-1}=\identity+\Energy/R$:
\[
 [\Op{B},\Filter]
 =\Filter(\Filter^{-1}\Op{B}-\Op{B}\Filter^{-1})\Filter
 =R^{-1}\Filter[\Energy,\Op{B}]\Filter.
\]
The scalar inequality
$u/(1+u/R)^2\le R/4$ for $u\ge0$ proves
\eqref{eq:averaged-energy} by spectral calculus.
\end{proof}

\subsection{Conditioned Boolean inputs and two-coordinate interactions}\label{sec:conditioned-law}

For independent coordinates, the conditional projections commute. After
conditioning, the order of resampling can matter.
Let $\Certificates$ be a family of nonempty subsets of $I$; its members
may have different sizes. Set
\begin{equation}\label{eq:conditioned-law}
 \begin{split}
 \Omega&\coloneqq
   \{\vx\in\{0,1\}^I:\vx_C\ne\vone_C
                           \text{ for every }C\in\Certificates\},\\
 \nu&\coloneqq\Bernoulli(p)^{\otimes I},\qquad
 \mu\coloneqq\nu(\,\cdot\mid\Omega),\qquad 0<p\le\tfrac12.
 \end{split}
\end{equation}
The set $\Omega$ is downward closed: changing ones to zeros keeps an input
in $\Omega$. It contains the all-zero input, so the conditioning is well
defined. In the rest of this section, all operators acting on the input
register, joint vectors, and infinity norms use $\mu$ unless
specified otherwise.

After all coordinates except $i$ are fixed, either setting $x_i=1$
completes a forbidden set and $i$ is forced to zero, or both values are
allowed and its conditional distribution is $\Bernoulli(p)$. In the latter case,
\eqref{eq:conditional-coordinates} applies with $q=p$. The conditional variance is then
\begin{equation}\label{eq:one-coordinate}
 (1-p)\norm{(\Deriv_i\vf)_0}^2+p\norm{(\Deriv_i\vf)_1}^2
       =p(1-p)\norm{\vf_1-\vf_0}^2.
\end{equation}
If $i$ is forced to zero, then $\Deriv_i=0$ on these inputs.
The squared norm $\norm{\Deriv_i\vf}^2$ averages this conditional variance over
the other coordinates.

Let $\Phase_e$ be the phase operator acting on the input register, as
defined in \eqref{eq:input-phase}, and put
\begin{equation}\label{eq:query-commutator}
 \Comm_e\coloneqq[\Deriv_e,\Phase_e].
\end{equation}
With all coordinates except $e$ fixed and both values of $x_e$ allowed,
write $\vf_0,\vf_1$ for the workspace vectors at $x_e=0,1$.
Equation~\eqref{eq:conditional-coordinates} also gives
\[
 (\Comm_e\vf)_0=2p\vf_1,\qquad
 (\Comm_e\vf)_1=-2(1-p)\vf_0.
\]
Taking the conditional squared norm gives
\[
 (1-p)\norm{(\Comm_e\vf)_0}^2+p\norm{(\Comm_e\vf)_1}^2
 =4p(1-p)\bigl((1-p)\norm{\vf_0}^2+p\norm{\vf_1}^2\bigr).
\]
If only one value of $x_e$ is allowed, $\Comm_e=0$ on that input. Hence
\begin{equation}\label{eq:same-coordinate}
 \norm{\Comm_e}\le2\sqrt p,
 \qquad \norm{\Comm_e\vf}_\infty\le2\norm\vf_\infty.
\end{equation}
If $i\ne e$, resampling coordinate $i$ leaves $x_e$ unchanged, so
$[\Deriv_i,\Phase_e]=0$. Thus $[\Energy,\Phase_e]=w_e\Comm_e$.
Applying \eqref{eq:resolvent-comm} with $\Op{B}=\Phase_e$ and using
$\norm{\Filter}\le1$ gives
\begin{equation}\label{eq:filter-Q}
 [\Phase_e,\Filter]=\frac{w_e}{R}\Filter\Comm_e\Filter,
 \qquad
 \norm{[\Phase_e,\Filter]}\le\frac{2w_e\sqrt p}{R}.
\end{equation}

\paragraph{Interactions between two coordinates.}\label{sec:two-coordinate}
We now quantify how two conditional projections
fail to commute. For distinct $i,j\in I$, define the scalar indicator
\begin{equation}\label{eq:completion-indicator}
 A_{ij}(\vx)\coloneqq
  \one_{\{\exists C\in\Certificates:\{i,j\}\subseteq C,
                  \ \vx_{C\setminus\{i,j\}}=\vone_{C\setminus\{i,j\}}\}}.
\end{equation}
Define the diagonal operator $\Op{A}_{ij}$ by
\[
 (\Op{A}_{ij}\vf)_{\vx}\coloneqq A_{ij}(\vx)\vf_{\vx}.
\]
Since $A_{ij}(\vx)$ depends only on coordinates outside $i,j$,
$[\Op{A}_{ij},\Deriv_i]=[\Op{A}_{ij},\Deriv_j]=0$. The event
$A_{ij}(\vx)=1$ contains every configuration on which resampling $i$ and $j$
can fail to commute.
The following lemma bounds these interactions.

\begin{lemma}[Two-coordinate interaction]\label[lemma]{lem:two-coordinate}
For distinct $i,j\in I$ and every $\vf\in\W\otimes\Input_\mu$,
\begin{equation}\label{eq:two-coordinate}
 \norm{[\Deriv_i,\Deriv_j]\vf}\le p\norm{\Op{A}_{ij}\vf},
\end{equation}
and
\begin{equation}\label{eq:two-coordinate-energy}
 \norm{[\Deriv_i,\Deriv_j]\vf}^2
 \le p^2(1+p)
       \left(\norm{\Op{A}_{ij}\Deriv_i\vf}^2
                      +\norm{\Op{A}_{ij}\Deriv_j\vf}^2\right).
\end{equation}
The projections also satisfy
\begin{equation}\label{eq:two-coordinate-order}
 \Deriv_i\Deriv_j+\Deriv_j\Deriv_i
       \succeq-p\Op{A}_{ij}(\Deriv_i+\Deriv_j).
\end{equation}
For every $\vx\in\Omega$,
\begin{equation}\label{eq:two-coordinate-coefficient}
 \norm{\bigl([\Deriv_i,\Deriv_j]\vf\bigr)_{\vx}}
       \le2pA_{ij}(\vx)\norm\vf_\infty.
\end{equation}
Both resampling orders preserve the coordinates outside $\{i,j\}$.
\end{lemma}

\begin{proof}
Fix an assignment to the coordinates outside $i,j$ that has positive
probability. Downward closure leaves one of the following allowed sets:
\[
 \{00\},\quad\{00,10\},\quad\{00,01\},\quad
 \{00,10,01\},\quad\{00,10,01,11\}.
\]
All except $\{00,10,01\}$ are products of one-coordinate sets, and their
conditional distributions factor. In these cases the two projections commute
and their product is positive semidefinite, which proves
\eqref{eq:two-coordinate}--\eqref{eq:two-coordinate-coefficient} in these cases.

In the remaining case, each single insertion is allowed but their
joint insertion is forbidden. There is therefore a set $C\in\Certificates$
containing both $i,j$, with $x_k=1$ for every $k\in C\setminus\{i,j\}$.
Thus $A_{ij}(\vx)=1$ on all
three inputs. The conditional probabilities of $00,10,01$ are
\[
 q_{00}\coloneqq\frac{1-p}{1+p},\qquad
 q_{10}\coloneqq\frac{p}{1+p},\qquad
 q_{01}\coloneqq\frac{p}{1+p}.
\]
The restrictions of $\Deriv_i$ and $\Deriv_j$ to these inputs are
$\Op{P}_i\coloneqq\ket{u_i}\bra{u_i}$ and
$\Op{P}_j\coloneqq\ket{u_j}\bra{u_j}$, respectively, where
\[
 \ket{u_i}\coloneqq\sqrt p\ket{00}-\sqrt{1-p}\ket{10},\qquad
 \ket{u_j}\coloneqq\sqrt p\ket{00}-\sqrt{1-p}\ket{01},\qquad
 \braket{u_i}{u_j}=p.
\]
Here $\vf_{ab}$ denotes the workspace vector at $x_i=a,x_j=b$,
with the outside assignment fixed. Suppressing those coordinates and
their common square root probability weight, the joint vector is
\[
 \sqrt{q_{00}}\vf_{00}\otimes\ket{00}
 +\sqrt{q_{10}}\vf_{10}\otimes\ket{10}
 +\sqrt{q_{01}}\vf_{01}\otimes\ket{01}.
\]
The projections act on its two input bits. Their sum has eigenvectors
$\ket{u_i}\pm\ket{u_j}$ with eigenvalues $1\pm p$, and
\[
 [\Op{P}_i,\Op{P}_j]
 =p\bigl(\ket{u_i}\bra{u_j}-\ket{u_j}\bra{u_i}\bigr).
\]
On $\mathscr V\coloneqq\operatorname{span}\{\ket{u_i},\ket{u_j}\}$,
this gives
\[
 [\Op{P}_i,\Op{P}_j]^\dagger[\Op{P}_i,\Op{P}_j]
       =p^2(1-p^2)\identity_{\mathscr V}
       \preceq p^2(1+p)(\Op{P}_i+\Op{P}_j)|_{\mathscr V}.
\]
Both projections vanish on $\mathscr V^\perp$. Also
\[
 \Op{P}_i\Op{P}_j+\Op{P}_j\Op{P}_i
       =(\Op{P}_i+\Op{P}_j)^2-(\Op{P}_i+\Op{P}_j)
       \succeq-p(\Op{P}_i+\Op{P}_j),
\]
by the same two eigenvalues. Averaging these bounds over the outside coordinates proves
\eqref{eq:two-coordinate}--\eqref{eq:two-coordinate-order}.

For \eqref{eq:two-coordinate-coefficient}, write
$\vg\coloneqq[\Deriv_i,\Deriv_j]\vf$ and compute using
\eqref{eq:conditional-coordinates}:
\[
 \vg_{00}=p^2(\vf_{10}-\vf_{01}),\qquad
 \vg_{10}=p(1-p)(\vf_{01}-\vf_{00}),\qquad
 \vg_{01}=p(1-p)(\vf_{00}-\vf_{10}).
\]
Each of $\vg_{00},\vg_{10},\vg_{01}$ has norm at most $2p\norm\vf_\infty$.
If $A_{ij}(\vx)=0$, the conditional distribution factors, so $\vg_{\vx}=0$.
Thus
\[
 \norm{\vg_{\vx}}\le2pA_{ij}(\vx)\norm\vf_\infty,
\]
which proves \eqref{eq:two-coordinate-coefficient}.
\end{proof}

\subsection{Conditional differences after averaging}\label{sec:coordinate-differences}

We derive two consequences of \cref{lem:two-coordinate}: a bound on sums of
averaged conditional variances over a common block of coordinates, and a
way to move a conditional difference through an averaging operator.

\paragraph{Summing over a common block.}
We seek a bound on $\sum_i\norm{\Filter_i\Deriv_i\vf}^2$ proportional
to $R\norm\vf^2$, even when many indices $i$ are summed. Bounding each term
separately by $\norm\vf^2$ would lose a factor equal to the number of
indices $i$. We instead look for a single averaging operator $\Filter$ over a common
set of coordinates such that $\Filter_i^2\preceq\Filter$ for every $i$.
This comparison gives
\[
 \sum_i\norm{\Filter_i\Deriv_i\vf}^2
 \le\sum_i\ip{\Deriv_i\vf}{\Filter\Deriv_i\vf},
\]
so all terms on the right use the same averaging operator.

To see what this common averaging operator accomplishes, suppose that the projections
$\Pi_i$ commute, and put $\Star\coloneqq\sum_i\Pi_i$ and
$\Filter\coloneqq(\identity+\Star/R)^{-1}$. Decompose $\vf$ into
orthogonal components on which every $\Pi_i$ acts either as the identity
or as zero. If a component $\vg$ is selected by exactly $k$ projections,
it appears in $k$ summands. At the same time, $\Star\vg=k\vg$, so
$\Filter\vg=R\vg/(R+k)$. Its total contribution is therefore
\[
 \sum_i\ip{\Pi_i\vg}{\Filter\Pi_i\vg}
 =\frac{kR}{R+k}\norm\vg^2
 \le R\norm\vg^2.
\]
The components counted more often are thus reduced more strongly by the
averaging operator. Summing over the orthogonal components gives a bound of
$R\norm\vf^2$, independent of the number of projections.

For dependent inputs, the conditional projections need not commute, so
this common decomposition is no longer available. The following lemma
controls the additional commutator terms in the sum.

\begin{lemma}[Summing projections]\label[lemma]{lem:projection-energy}
Let $\Pi_1,\ldots,\Pi_\ell$ be orthogonal projections on a
finite-dimensional Hilbert space, and put
$\Star\coloneqq\sum_{i=1}^{\ell}\Pi_i$.
Suppose that $\gamma>0$ and $\kappa\ge0$ satisfy
\begin{equation}\label{eq:projection-energy-hypotheses}
 \Star^2\succeq\gamma\Star,
 \qquad
 \sum_{i=1}^{\ell}\norm{[\Pi_i,\Star]\vg}^2
       \le\kappa^2\ip\vg{\Star\vg}
       \quad\text{for every $\vg$}.
\end{equation}
For every $R>0$, let
$\Filter\coloneqq(\identity+\Star/R)^{-1}$. Then
\begin{equation}\label{eq:projection-energy}
 \sum_{i=1}^{\ell}\ip{\Pi_i\vf}{\Filter\Pi_i\vf}
 \le\left(2+\frac{\kappa^2}{\gamma^2}\right)R\norm\vf^2
 \quad\text{for every $\vf$}.
\end{equation}
\end{lemma}

\begin{proof}
Put
$Y\coloneqq\sum_i\ip{\Pi_i\vf}{\Filter\Pi_i\vf}$.
Write $\Filter\Pi_i=\Pi_i\Filter+[\Filter,\Pi_i]$ and use
$\Pi_i^2=\Pi_i$. Applying \eqref{eq:resolvent-comm} gives
\begin{equation}\label{eq:projection-energy-identity}
 \begin{aligned}
 Y&=\operatorname{Re}\sum_i
       \ip{\Pi_i\vf}{\Pi_i\Filter\vf+[\Filter,\Pi_i]\vf}\\
  &=\ip\vf{\Star\Filter\vf}
    +\frac1R\operatorname{Re}\sum_i
       \ip{\Filter\Pi_i\vf}{[\Pi_i,\Star]\Filter\vf}.
 \end{aligned}
\end{equation}
The first term is at most $R\norm\vf^2$, since
$\Star\Filter=R(\identity-\Filter)$ and
$0\preceq\Filter\preceq\identity$.

The first hypothesis in \eqref{eq:projection-energy-hypotheses} says that
every positive eigenvalue of $\Star$ is at least $\gamma$. For
$\vh\in\ker\Star$,
\[
 0=\ip\vh{\Star\vh}=\sum_i\norm{\Pi_i\vh}^2,
\]
so every $\Pi_i$ annihilates $\vh$. By self-adjointness,
$\Pi_i\vf$ is therefore orthogonal to $\ker\Star$.
On this orthogonal complement, the eigenvalues of $\Filter$ are at most
$R/(R+\gamma)$, so $\Filter^2\preceq R\Filter/(R+\gamma)$ there.
It follows that
\begin{equation}\label{eq:projection-energy-spectral}
 \sum_i\norm{\Filter\Pi_i\vf}^2
       \le\frac R{R+\gamma}Y,
 \qquad
 \ip{\Filter\vf}{\Star\Filter\vf}
       \le\frac{R^2}{R+\gamma}\norm\vf^2.
\end{equation}
For the second inequality in \eqref{eq:projection-energy-spectral},
a positive eigenvalue $\lambda\ge\gamma$
contributes the factor
$\lambda R^2/(R+\lambda)^2\le R^2/(R+\gamma)$, while a zero eigenvalue
contributes zero.

Apply Cauchy--Schwarz to the sum in
\eqref{eq:projection-energy-identity}, and then use the second hypothesis
in \eqref{eq:projection-energy-hypotheses} with $\vg=\Filter\vf$.
Together with \eqref{eq:projection-energy-spectral}, this gives
\[
 \begin{aligned}
 Y&\le R\norm\vf^2+
       \frac\kappa R
       \left(\sum_i\norm{\Filter\Pi_i\vf}^2\right)^{1/2}
       \ip{\Filter\vf}{\Star\Filter\vf}^{1/2}\\
  &\le R\norm\vf^2+
       \frac{\kappa\sqrt R}{R+\gamma}\sqrt Y\norm\vf.
 \end{aligned}
\]
If $\vf=\vzero$, the conclusion is immediate. Otherwise set
$y\coloneqq Y/(R\norm\vf^2)$. Since $R+\gamma\ge\gamma$,
\[
 y\le1+\frac\kappa{R+\gamma}\sqrt y
   \le1+\frac y2+\frac{\kappa^2}{2\gamma^2}.
\]
Rearranging proves \eqref{eq:projection-energy}.
\end{proof}

We now return to the conditioned Bernoulli distribution $\mu$ defined in
\eqref{eq:conditioned-law} and its projections $\Deriv_i$.
We call a set $B\subseteq I$ a block of coordinates. After
fixing all coordinates outside $B$, the conditional operations within
$B$ act on a single conditional input distribution. If each coordinate
can interact with only a few others throughout this conditional
distribution, \cref{lem:projection-energy} gives the desired bound. The next
statement also allows a small probability of outside assignments for
which this condition fails.

\begin{lemma}[A common block of coordinates]\label[lemma]{lem:common-set}
Let $B\subseteq I$, and define
\[
 \Star_B\coloneqq\sum_{i\in B}\Deriv_i,
 \qquad
 N_{i,B}(\vx)\coloneqq
       \sum_{j\in B\setminus\{i\}}A_{ij}(\vx)
       \quad(i\in B).
\]
Suppose that $\mathcal G_B$ is an event determined by the coordinates
outside $B$, with $\mu(\mathcal G_B^c)\le b$, where $0\le b\le1$.
Assume that, for some $d\ge0$ with $pd\le1/10$,
\begin{equation}\label{eq:common-set-hypothesis}
 N_{i,B}(\vx)\le d
 \quad\text{for every $i\in B$ and
 every $\vx\in\supp\mu\cap\mathcal G_B$}.
\end{equation}
For each $i\in B$, let $\Energy_i$ be a positive semidefinite operator
acting on the input register and satisfying $\Energy_i\succeq\Star_B$.
For any $R>0$, put
\[
 \Filter_B\coloneqq(\identity+\Star_B/R)^{-1},
 \qquad
 \Filter_i\coloneqq(\identity+\Energy_i/R)^{-1}.
\]
Then every joint vector $\vf\in\W\otimes\Input_\mu$ satisfies
\begin{equation}\label{eq:common-set-bound}
 \sum_{i\in B}\norm{\Filter_i\Deriv_i\vf}^2
 \le\sum_{i\in B}\ip{\Deriv_i\vf}{\Filter_B\Deriv_i\vf}
 \le3R\norm\vf^2+|B|b\norm\vf_\infty^2.
\end{equation}
The infinity norm uses $\mu$.
\end{lemma}

\begin{proof}
The assertion is immediate if $B=\varnothing$. By \eqref{eq:filter-order},
\[
 \Filter_i^2\preceq\Filter_i\preceq\Filter_B
       \quad(i\in B),
\]
which proves the first inequality in \eqref{eq:common-set-bound}.

First suppose that \eqref{eq:common-set-hypothesis} holds throughout
$\supp\mu$. By \eqref{eq:two-coordinate-order},
\[
 \Star_B^2
 \succeq\Star_B-p\sum_{i\in B}\sum_{j\in B\setminus\{i\}}
                         \Op{A}_{ij}\Deriv_i
 \succeq(1-pd)\Star_B.
\]
For the last inequality, each $\Op{A}_{ij}$ commutes with $\Deriv_i$.
Since $\sum_{j\in B\setminus\{i\}}A_{ij}(\vx)=N_{i,B}(\vx)\le d$,
the operator $\sum_{j\in B\setminus\{i\}}\Op{A}_{ij}$ lies between
$0$ and $d\identity$, and hence
\[
 0\preceq\sum_{j\in B\setminus\{i\}}\Op{A}_{ij}\Deriv_i
       \preceq d\Deriv_i.
\]

For each $i\in B$ and input $\vx\in\supp\mu$, at most $d$ of the
workspace vectors
\[
 ([\Deriv_i,\Deriv_j]\vg)_{\vx},
 \qquad j\in B\setminus\{i\},
\]
can be nonzero. Applying Cauchy--Schwarz for each $\vx$, averaging over
$\vx\sim\mu$, and then using \eqref{eq:two-coordinate-energy} yields
\begin{align*}
 \sum_{i\in B}\norm{[\Deriv_i,\Star_B]\vg}^2
 &\le d\sum_{i\in B}\sum_{j\in B\setminus\{i\}}
                      \norm{[\Deriv_i,\Deriv_j]\vg}^2\\
 &\le dp^2(1+p)\sum_{i\in B}\sum_{j\in B\setminus\{i\}}
       \left(\norm{\Op{A}_{ij}\Deriv_i\vg}^2
             +\norm{\Op{A}_{ij}\Deriv_j\vg}^2\right)\\
 &=2dp^2(1+p)\sum_{i\in B}
       \E_{\vx\sim\mu}
       \left[N_{i,B}(\vx)\norm{(\Deriv_i\vg)_{\vx}}^2\right]\\
 &\le2(1+p)(pd)^2\ip\vg{\Star_B\vg}.
\end{align*}
The equality uses $A_{ij}(\vx)=A_{ji}(\vx)$ and
$A_{ij}(\vx)^2=A_{ij}(\vx)$.
Thus \cref{lem:projection-energy} applies with
\[
 \gamma\coloneqq1-pd\ge\frac9{10},
 \qquad
 \kappa^2\coloneqq2(1+p)(pd)^2\le\frac3{100}.
\]
Since $2+\kappa^2/\gamma^2<3$, it follows that
\begin{equation}\label{eq:common-set-uniform}
 \sum_{i\in B}\ip{\Deriv_i\vf}{\Filter_B\Deriv_i\vf}
       \le3R\norm\vf^2.
\end{equation}

For the general case, fix the coordinates outside $B$. Each
$\Deriv_i$, $i\in B$, and hence $\Filter_B$, preserves these outside
coordinates. Within the resulting conditional input distribution, the
joint vector has the original workspace vectors and the conditional
square root probability weights. The distribution is still obtained by
conditioning independent Bernoulli bits on a downward-closed set of allowed
assignments within $B$, so
\eqref{eq:two-coordinate}--\eqref{eq:two-coordinate-coefficient} remain valid there.
On an outside assignment in $\mathcal G_B$, \eqref{eq:common-set-hypothesis}
holds for every permitted assignment within $B$, so the preceding
argument proves \eqref{eq:common-set-uniform} in the conditional norm.

On any other outside assignment, $0\preceq\Filter_B\preceq\identity$
and the fact that the $\Deriv_i$ are projections give, in the conditional norm,
\[
 \sum_{i\in B}\ip{\Deriv_i\vf}{\Filter_B\Deriv_i\vf}
 \le\sum_{i\in B}\norm{\Deriv_i\vf}^2
 \le |B|\norm\vf^2.
\]
Averaging the good conditional bounds contributes at most
$3R\norm\vf^2$. Averaging the remaining bounds contributes at most
$|B|\mu(\mathcal G_B^c)\norm\vf_\infty^2$.
This proves the second inequality in \eqref{eq:common-set-bound}.
\end{proof}

\paragraph{Moving a difference through an averaging operator.}\label{sec:moving}
\Cref{lem:common-set} controls sums of
$\norm{\Filter\Deriv_i\vf}^2$. Query errors instead produce the opposite order,
$\Deriv_i\Filter\vf$. The following lemma compares the two orders.

\begin{lemma}[Moving a difference through an averaging operator]
\label[lemma]{lem:difference-average}
Let $\Energy=\sum_jw_j\Deriv_j$, where $w_j\ge0$ and
$\sum_jw_j\le w$. Fix $i\in I$. Suppose an event $\mathcal G$ satisfies
$\mu(\mathcal G^c)\le b$, and
\[
 \sum_{j\ne i}w_jA_{ij}(\vx)\le d\quad(\vx\in\mathcal G),
 \qquad 0\le d\le w,\qquad pd\le\frac25.
\]
For $R\ge1$ and $\Filter=(\identity+\Energy/R)^{-1}$, every joint vector
$\vf\in\W\otimes\Input_\mu$ satisfies
\begin{equation}\label{eq:difference-average}
 \norm{\Deriv_i\Filter\vf}^2
 \le4\norm{\Filter\Deriv_i\vf}^2
       +\frac pR\norm\vf^2
       +\frac{64p^2w^2b}{R^2}\norm\vf_\infty^2.
\end{equation}
Here $\mathcal G$ may depend on all input coordinates.
\end{lemma}

\begin{proof}
Put $\tau^2\coloneqq16p^2w^2b$. We first prove
\begin{equation}\label{eq:weighted-interaction}
 \norm{[\Deriv_i,\Energy]\vf}^2
 \le(1+p)(pd)^2\norm{\Deriv_i\vf}^2
       +(1+p)p^2d\,\ip\vf{\Energy\vf}
                        +\tau^2\norm\vf_\infty^2.
\end{equation}
For $j\ne i$, \eqref{eq:two-coordinate-coefficient} gives
$([\Deriv_i,\Deriv_j]\vf)_{\vx}=0$ whenever $A_{ij}(\vx)=0$.
For each input $\vx\in\supp\mu\cap\mathcal G$, weighted
Cauchy--Schwarz therefore gives
\[
 \norm{\Bigl(\sum_{j\ne i}w_j[\Deriv_i,\Deriv_j]\vf\Bigr)_{\vx}}^2
       \le d\sum_{j\ne i}w_j
                     \norm{([\Deriv_i,\Deriv_j]\vf)_{\vx}}^2.
\]
Multiplying by $\mu(\vx)$ and summing over
$\vx\in\supp\mu\cap\mathcal G$, then extending the nonnegative sum to
all $\vx\in\supp\mu$ and applying \eqref{eq:two-coordinate-energy},
bounds this contribution by
\[
 dp^2(1+p)\sum_{j\ne i}w_j
       \left(\norm{\Op{A}_{ij}\Deriv_i\vf}^2
                     +\norm{\Op{A}_{ij}\Deriv_j\vf}^2\right).
\]
For the first sum, split the expectation according to whether
$\vx\in\mathcal G$. On $\mathcal G$, use
$\sum_{j\ne i}w_jA_{ij}(\vx)\le d$; on its complement, use
$\sum_{j\ne i}w_jA_{ij}(\vx)\le w$ and
$\norm{\Deriv_i\vf}_\infty\le2\norm\vf_\infty$. This gives
\[
 \sum_{j\ne i}w_j\norm{\Op{A}_{ij}\Deriv_i\vf}^2
 =\E_{\vx\sim\mu}\left[
      \left(\sum_{j\ne i}w_jA_{ij}(\vx)\right)
      \norm{(\Deriv_i\vf)_{\vx}}^2\right]
 \le d\norm{\Deriv_i\vf}^2+4wb\norm\vf_\infty^2.
\]
The second sum is at most
\[
 \sum_jw_j\norm{\Deriv_j\vf}^2=\ip\vf{\Energy\vf}.
\]

For $\vx\in\supp\mu\cap\mathcal G^c$, \eqref{eq:two-coordinate-coefficient} gives
\[
 \norm{([\Deriv_i,\Energy]\vf)_{\vx}}
       \le2pw\norm\vf_\infty.
\]
The inputs in $\mathcal G^c$ contribute at most
$4p^2w^2b\norm\vf_\infty^2$.
Together, the two exceptional contributions have coefficient at most
\[
 \bigl(4(1+p)p^2dw+4p^2w^2\bigr)b
       \le16p^2w^2b=\tau^2.
\]
This proves \eqref{eq:weighted-interaction}. Both uses of $\mathcal G$
split an expectation according to the input $\vx$ indexing the workspace
vector: first $([\Deriv_i,\Energy]\vf)_{\vx}$, then
$(\Deriv_i\vf)_{\vx}$. The event can therefore depend on all coordinates, as
allowed in the statement.

Apply \eqref{eq:resolvent-comm} and
\eqref{eq:weighted-interaction} to $\Filter\vf$. Using
$\norm{\Filter\vf}\le\norm\vf$ and
$\norm{\Filter\vf}_\infty\le\norm\vf_\infty$, together with
\eqref{eq:averaged-energy}, gives
\begin{equation}\label{eq:relative-average-commutator}
 \norm{[\Deriv_i,\Filter]\vf}^2
 \le\frac{(1+p)(pd)^2}{R^2}
                    \norm{\Deriv_i\Filter\vf}^2
       +\frac{(1+p)p^2d}{4R}\norm\vf^2
                   +\frac{\tau^2}{R^2}\norm\vf_\infty^2.
\end{equation}
Now $pd\le2/5$ and $p\le1/2$ imply
$(1+p)(pd)^2\le6/25<1/4$ and $(1+p)pd\le3/5<1$.
Using
$\Deriv_i\Filter\vf=\Filter\Deriv_i\vf+[\Deriv_i,\Filter]\vf$
and $R\ge1$, we obtain
\[
 \norm{\Deriv_i\Filter\vf}^2
 \le2\norm{\Filter\Deriv_i\vf}^2
       +\frac12\norm{\Deriv_i\Filter\vf}^2
       +\frac p{2R}\norm\vf^2
       +\frac{2\tau^2}{R^2}\norm\vf_\infty^2.
\]
Move the second term to the left and substitute the definition of
$\tau^2$ to prove \eqref{eq:difference-average}. The smallness condition
on $pd$ is used only in this last absorption step.
\end{proof}

%% file: sections/triangle.tex
\section{Triangle finding}\label{sec:triangle}\label{sec:fixed-graph}

We first prove a bound in terms of the number of triangles and the maximum
degree of a fixed graph, then apply it to $K_{n,n,n}$. All constants in
this section are absolute.

Let $G$ be a finite simple undirected graph known to the algorithm. The
input is an unknown subgraph $G_{\vx}$ on the same vertex set, specified by
$\vx\in\{0,1\}^I$, where $I\coloneqq E(G)$ and $x_e=1$ precisely when $e$
is present. Pairs outside $I$, including repeated vertices, have known answer
zero. Write $m\coloneqq|I|$, let $\Delta$ be the maximum degree of $G$,
and let $\Triangles$ be its family of triangles, each identified with its
three-edge set. Put $M\coloneqq|\Triangles|$, and assume $M\ge1$.
Thus $m\ge3$ and $\Delta\ge2$.

For the positive input distribution, choose $T\in\Triangles$ uniformly,
set its edges to one, and sample the other edges independently with
probability $p$, conditioned on $T$ being the unique triangle. This
conditioning has positive probability, since all edges outside $T$ can be
zero. Denote the distribution for a fixed $T$ by $\mu_T$, and put
$\mu_+\coloneqq M^{-1}\sum_T\mu_T$. The algorithm is given $G$ but is not
told $T$. Write $\sigma$ for its average probability of finding the triangle.

\begin{theorem}[Triangle search in a fixed graph]\label[theorem]{thm:fixed-graph}
There are absolute constants $C_0,C_1>0$ such that the following holds.
Let $G$ have $m$ edges, maximum degree $\Delta$, and $M\ge1$ triangles. If
\begin{equation}\label{eq:density-range}
 0<p\le\frac1{10\sqrt\Delta},\qquad \Delta p\ge C_0\log(2m),
\end{equation}
then every $t$-query quantum algorithm has average success probability $\sigma$
under $\mu_+$ satisfying
\begin{equation}\label{eq:fixed-graph-bound}
 \sqrt\sigma\le C_1(t+3)\frac{\Delta^{1/3}}{\sqrt M\,p^{1/6}}.
\end{equation}
\end{theorem}

We first prove \cref{thm:fixed-graph}, assuming \eqref{eq:density-range}
throughout and taking $C_0$ large enough that $\Delta p\ge1$. We then
specialize to the complete tripartite graph.

\subsection{Input distributions and local comparison}\label{sec:triangle-comparison}
Let $\Omega_-$ be the set of triangle-free subgraphs of $G$, and put
\[
 \nu\coloneqq\Bernoulli(p)^{\otimes I},\qquad
 \mu_-\coloneqq\nu(\,\cdot\mid\Omega_-),\qquad
 \X\coloneqq\Input_{\mu_-}.
\]
The all-zero input makes this conditioning well defined. From now on,
the conditional averages $\Cond_i$ and differences $\Deriv_i$ use this
negative distribution. It is an instance of the conditioned Bernoulli distribution in
\cref{sec:conditioned-law}, with forbidden sets $\Triangles$. Unless another
distribution is displayed, $\vf$ denotes a vector in $\W\otimes\X$, and
$\norm\vf_\infty=\max_{\vx\in\Omega_-}\norm{\vf_{\vx}}$.

For $T\in\Triangles$, write $\bar T\coloneqq I\setminus T$ and
$\nu_{\bar T}\coloneqq\Bernoulli(p)^{\otimes\bar T}$. Let
$\mathcal G_T$ be the event on these outside coordinates that setting all
three edges of $T$ to one gives a graph whose unique triangle is $T$.
Thus $\mu_T$ samples $\vz$ from
$\nu_{\bar T}(\,\cdot\mid\mathcal G_T)$ and returns $(\vz,\vone_T)$.
We retain $T$ as a label in the positive input register:
\[
 \Y_T\coloneqq\Input_{\mu_T},\qquad
 \Y\coloneqq\operatorname{span}
       \{\ket{T,\vy}:T\in\Triangles,\ \vy\in\supp\mu_T\}.
\]
A positive joint vector has the form
\begin{equation}\label{eq:positive-space}
 \begin{gathered}
 \vg=\frac1{\sqrt M}\sum_{T\in\Triangles}
       \sum_{\vy\in\supp\mu_T}
          \sqrt{\mu_T(\vy)}\,\vg_{T,\vy}\otimes\ket{T,\vy},\\
 \vg_T\coloneqq\sum_{\vy\in\supp\mu_T}
          \sqrt{\mu_T(\vy)}\,\vg_{T,\vy}\otimes\ket\vy,
 \qquad \norm\vg^2=\frac1M\sum_T\norm{\vg_T}^2.
 \end{gathered}
\end{equation}
The component $\vg_T\in\W\otimes\Y_T$ includes the weights
$\sqrt{\mu_T(\vy)}$, but excludes $1/\sqrt M$.
The normalized input states are
\[
 \begin{gathered}
 \ket{\mu_-}\coloneqq\sum_{\vx\in\Omega_-}
             \sqrt{\mu_-(\vx)}\ket\vx,\\
 \ket{\mu_+}\coloneqq\frac1{\sqrt M}
       \sum_T\sum_{\vy\in\supp\mu_T}\sqrt{\mu_T(\vy)}\ket{T,\vy}.
 \end{gathered}
\]
The label is determined by the input's unique triangle and is not
accessible to the algorithm. The phase operators $\Phase_e^-$ and
$\Phase_e^+$ act on the negative and positive input registers,
respectively, as defined in
\eqref{eq:input-phase}. The same notation $\Phase_e^+$ on $\Y_T$
denotes its restriction to that component.

We will compare workspace vectors at positive inputs with those at a
few negative inputs. To compare their probability weights, we first bound increasing
events under $\mu_-$ by their probabilities under the product distribution $\nu$.

\begin{lemma}[Comparison with independent bits]\label[lemma]{lem:domination}
Let $\Omega\subseteq\{0,1\}^I$ be nonempty and downward closed, and put
$\mu\coloneqq\nu(\,\cdot\mid\Omega)$, where
$\nu=\Bernoulli(p)^{\otimes I}$.
There is a coupling $(\vX,\vY)$ with $\vX\sim\mu$, $\vY\sim\nu$, and
$X_i\le Y_i$ for every $i\in I$. Consequently, every increasing event
has probability under $\mu$ at most its probability under $\nu$.
\end{lemma}
\begin{proof}
We construct both inputs using the same independent random numbers. The
key point is that, even after some coordinates have been chosen, the
conditional probability of a one under $\mu$ is at most $p$.
To see this, let $\boldsymbol Z\sim\mu$ and fix all coordinates except $i$
to an assignment of positive probability. Downward closure leaves either
only $Z_i=0$ or both values allowed. In the latter case their conditional
probabilities are $1-p$ and $p$, respectively. For any
$J\subseteq I\setminus\{i\}$, averaging
over the coordinates outside $J\cup\{i\}$ therefore gives
\begin{equation}\label{eq:domination-marginal}
 \Pr[Z_i=1\mid\boldsymbol Z_J]
 =\E\!\left[
     \Pr[Z_i=1\mid\boldsymbol Z_{I\setminus\{i\}}]
       \,\middle|\,\boldsymbol Z_J\right]
 \le p.
\end{equation}

Relabel the coordinates as $I=[m]$. For each prefix
$(x_1,\ldots,x_{i-1})$ of positive probability under $\mu$, put
\[
 q_i(x_1,\ldots,x_{i-1})\coloneqq
 \Pr[Z_i=1\mid Z_1=x_1,\ldots,Z_{i-1}=x_{i-1}].
\]
Equation~\eqref{eq:domination-marginal} gives $q_i\le p$. Sample $U_1,\ldots,U_m$ independently
and uniformly from $(0,1)$, and define successively
\[
 X_i\coloneqq\one_{\{U_i\le q_i(X_1,\ldots,X_{i-1})\}},
 \qquad
 Y_i\coloneqq\one_{\{U_i\le p\}}.
\]
For $i=1$, the conditioning in $q_1$ is empty. At step $i$, the variable
$U_i$ is independent of $X_1,\ldots,X_{i-1}$, so the conditional
probability that $X_i=1$ is exactly
$q_i(X_1,\ldots,X_{i-1})$. Induction on $i$ shows that every prefix of
$\vX$ has the corresponding marginal distribution under $\mu$, and hence
$\vX\sim\mu$. Each $Y_i$ depends only on $U_i$ and is
$\Bernoulli(p)$, so the $Y_i$ are independent and $\vY\sim\nu$.

Since $q_i\le p$, whenever $X_i=1$ the same random number also gives
$Y_i=1$. Thus $X_i\le Y_i$ for every coordinate. If $\mathcal A$ is an
increasing event, $\vX\in\mathcal A$ then implies $\vY\in\mathcal A$,
and consequently
\[
 \mu(\mathcal A)=\Pr[\vX\in\mathcal A]
 \le\Pr[\vY\in\mathcal A]=\nu(\mathcal A).\qedhere
\]
\end{proof}

Starting from a negative input, filling $T$ can create another triangle
only if that triangle shares an edge with $T$. Its two other edges must
already be present. Distinct triangles share at most one edge, and there
are at most $\Delta$ triangles through any fixed edge. Applying
\cref{lem:domination} to $\mu_-$ and taking a union bound therefore give
\begin{equation}\label{eq:G-probability}
 \mu_-(\mathcal G_T)\ge1-3\Delta p^2\ge0.97.
\end{equation}
Here $\mathcal G_T$ is evaluated on a full input by looking only at its
edges outside $T$.

Put $Z\coloneqq\nu(\Omega_-)$ and
$Z_T\coloneqq\nu_{\bar T}(\mathcal G_T)$. On $\mathcal G_T$, exactly
the seven proper subsets of $T$ give negative inputs. Hence
\begin{equation}\label{eq:partition-ratio}
 \mu_-(\mathcal G_T)=\frac{(1-p^3)Z_T}{Z},\qquad
 r_T\coloneqq\frac Z{(1-p)^3Z_T}
       =\frac{1-p^3}{(1-p)^3\mu_-(\mathcal G_T)}\le c.
\end{equation}
The last inequality uses $p\le1/2$ and \eqref{eq:G-probability}.
We will use the bound $r_T\le c$ later in the proof.

\paragraph{A comparison using four negative inputs.}
For $\vy\in\supp\mu_T$ and $U\subseteq T$, let $\vy^U$ agree with
$\vy$ outside $T$ and have exactly the edges in $U$ present inside $T$.
Every proper $U$ gives a negative input: deletion cannot create a different
triangle, and at least one edge of $T$ is missing.

As explained in \cref{sec:overview-local}, we combine the workspace vectors at the
four most probable of these inputs, $\vy^\varnothing$ and $\vy^{\{i\}}$ for $i\in T$.
We seek a linear map $\Local_T:\X\to\Y_T$ satisfying
$\Local_T\ket{\mu_-}=\ket{\mu_T}$ and giving the same result whether an edge of $T$
is queried before or after applying the map, starting from $\ket{\mu_-}$.
These two requirements determine its scalar coefficients:
\begin{equation}\label{eq:local-transfer}
 (\Local_T\vf)_{\vy}\coloneqq
       \sum_{i\in T}\vf_{\vy^{\{i\}}}-2\vf_{\vy^\varnothing}.
\end{equation}
The resulting joint vector is
$\sum_{\vy}\sqrt{\mu_T(\vy)}(\Local_T\vf)_{\vy}\otimes\ket\vy$.
Thus \eqref{eq:local-transfer} combines workspace vectors, while the
probability weights in the resulting vector are those of $\mu_T$. For every $e\in T$,
\[
 \Local_T\Phase_e^-\ket{\mu_-}=-\ket{\mu_T}
 =\Phase_e^+\Local_T\ket{\mu_-}.
\]

For brevity, define the positive joint vectors
\[
 \vf_0^{(T)}\coloneqq\sum_{\vy}\sqrt{\mu_T(\vy)}
           \vf_{\vy^\varnothing}\otimes\ket\vy,
 \qquad
 \vf_i^{(T)}\coloneqq\sum_{\vy}\sqrt{\mu_T(\vy)}
           \vf_{\vy^{\{i\}}}\otimes\ket\vy\quad(i\in T),
\]
where both sums range over $\supp\mu_T$.
For each $\vz\in\mathcal G_T$, \eqref{eq:partition-ratio} gives
\begin{equation}\label{eq:triangle-input-weights}
 \mu_T(\vz,\vone_T)
 =\frac{\nu_{\bar T}(\vz)}{Z_T}
 =r_T\mu_-(\vz,\vzero_T).
\end{equation}
We will use the following two consequences:
\begin{equation}\label{eq:comparison}
 \norm{\vf_0^{(T)}}^2\le c\norm\vf^2,
 \qquad
 \norm{\vf_i^{(T)}-\vf_0^{(T)}}^2
       \le\frac cp\norm{\Deriv_i\vf}^2.
\end{equation}
Multiplying \eqref{eq:triangle-input-weights} by $\norm{\vf_{(\vz,\vzero_T)}}^2$ and
summing over $\vz\in\mathcal G_T$ proves the first inequality in
\eqref{eq:comparison}, using $r_T\le c$ from \eqref{eq:partition-ratio}.
For the second inequality in \eqref{eq:comparison}, fix
$\vz\in\mathcal G_T$ and set the other two
edges of $T$ to zero. Both values of $x_i$ are allowed, so by
\eqref{eq:one-coordinate} these two negative inputs contribute
\[
 \frac{p(1-p)^3\nu_{\bar T}(\vz)}Z
       \norm{\vf_{\vy^{\{i\}}}-\vf_{\vy^\varnothing}}^2,
 \qquad \vy\coloneqq(\vz,\vone_T),
\]
to $\norm{\Deriv_i\vf}^2$. Here $(1-p)^2\nu_{\bar T}(\vz)/Z$ is the
total probability of the two inputs, and $p(1-p)$ is their
conditional variance factor.
Summing over the distinct assignments $\vz\in\mathcal G_T$ yields
$\norm{\Deriv_i\vf}^2\ge(p/r_T)
\norm{\vf_i^{(T)}-\vf_0^{(T)}}^2$, proving the bound.

\begin{lemma}[Local comparison]\label[lemma]{lem:local-transfer}
We have $\Local_T\ket{\mu_-}=\ket{\mu_T}$. For every $\vf$,
\begin{align}
 \norm{\Local_T\vf}^2
 &\le c\left(\norm\vf^2+\frac1p
                 \sum_{i\in T}\norm{\Deriv_i\vf}^2\right),
       \label{eq:transfer-energy}\\
 \norm{\Local_T\vf}^2&\le\frac cp\norm\vf^2.
       \label{eq:transfer-rough}
\end{align}
For $e\notin T$, $\Phase_e^+\Local_T=\Local_T\Phase_e^-$.
For $e\in T$,
\begin{align}
 (\Phase_e^+\Local_T-\Local_T\Phase_e^-)\vf
 &=-2\sum_{i\in T\setminus\{e\}}
                  (\vf_i^{(T)}-\vf_0^{(T)}),
       \label{eq:transfer-exact}\\
 \norm{(\Phase_e^+\Local_T-\Local_T\Phase_e^-)\vf}^2
 &\le\frac cp\sum_{i\in T\setminus\{e\}}
                   \norm{\Deriv_i\vf}^2.
       \label{eq:transfer-defect}
\end{align}
\end{lemma}
\begin{proof}
The four scalar coefficients in \eqref{eq:local-transfer} sum to one, so
$\Local_T\ket{\mu_-}=\ket{\mu_T}$.
Write $\Local_T\vf=\vf_0^{(T)}+
\sum_{i\in T}(\vf_i^{(T)}-\vf_0^{(T)})$ and apply
\eqref{eq:comparison} to its four terms. This proves
\eqref{eq:transfer-energy}, and contraction of each $\Deriv_i$ gives
\eqref{eq:transfer-rough}. For $e\notin T$, all four inputs have the same
value of $x_e$, so $\Phase_e^+\Local_T=\Local_T\Phase_e^-$. For $e\in T$,
the query gives phase $-1$ on $\vy$ and
$\vy^{\{e\}}$, and phase $+1$ on the other three evaluated inputs.
Substituting these phases gives \eqref{eq:transfer-exact}; then
\eqref{eq:comparison} gives \eqref{eq:transfer-defect}.
\end{proof}

\subsection{Star averages and sums over triangles}\label{sec:triangle-energy}
We now define the averages of \cref{sec:overview-smoothing} for the triangle-free
distribution. For each vertex $w$, put
\[
 \st(w)\coloneqq\{i\in I:w\in i\},\qquad
 \Star_w\coloneqq\sum_{i\in\st(w)}\Deriv_i.
\]
For a joint vector $\vf$, the sum of conditional variances over the edges
incident to $w$ is
\[
 \sum_{i\in\st(w)}\norm{\Deriv_i\vf}^2
       =\ip\vf{\Star_w\vf}.
\]
For $R>0$ and $T\in\Triangles$, define
\begin{equation}\label{eq:triangle-averages}
 \begin{gathered}
 \Energy_T\coloneqq\sum_{w\in V(T)}\Star_w
       =\sum_{i\in I}h_T(i)\Deriv_i,\qquad
 h_T(i)\coloneqq|i\cap V(T)|,\\
 \Filter_w\coloneqq(\identity+\Star_w/R)^{-1},\qquad
 \Filter_T\coloneqq(\identity+\Energy_T/R)^{-1}.
 \end{gathered}
\end{equation}
The weights satisfy $0\le h_T(i)\le2$ and $\sum_i h_T(i)\le3\Delta$.
By \cref{lem:filter}, $\Filter_w\ket{\mu_-}=\Filter_T\ket{\mu_-}=\ket{\mu_-}$,
and both averaging operators contract $\norm{\cdot}$ and $\norm{\cdot}_\infty$.
All triangles containing $w$ can be
compared with the same averaging operator $\Filter_w$: since $\Energy_T\succeq\Star_w$
for $w\in V(T)$, \eqref{eq:filter-order} gives
\begin{equation}\label{eq:triangle-average-order}
 \Filter_T^2\preceq\Filter_T\preceq\Filter_w.
\end{equation}
To check the local hypotheses for these averages, we first control degrees
in the random input. Define
\[
 \deg_w(\vx)\coloneqq\sum_{i\in\st(w)}x_i,
 \qquad \deg_{\max}(\vx)\coloneqq\max_w\deg_w(\vx),
 \qquad d_0\coloneqq\lceil8\Delta p\rceil.
\]
Thus $\deg_w$ counts present edges in the input, whereas $|\st(w)|\le\Delta$
counts edges of the fixed graph. Since $\Delta p\ge1$,
\begin{equation}\label{eq:triangle-cutoff}
 d_0\le9\Delta p,\qquad pd_0\le9\Delta p^2\le\frac9{100}.
\end{equation}
Under the product distribution, an exponential moment bound gives
\[
 \Pr_\nu(\deg_w\ge8\Delta p)
 \le e^{-8\Delta p\log8}(1+7p)^{\deg_G(w)}
 \le\exp\bigl(-(8\log8-7)\Delta p\bigr).
\]
There are at most $2m$ nonisolated vertices. By \cref{lem:domination},
the degree event at each vertex is no more likely under $\mu_-$ than under
the product distribution. A union bound therefore gives
\begin{equation}\label{eq:triangle-tail}
 \Pr_{\mu_-}(\deg_{\max}>d_0)\le b,
 \qquad b\coloneqq\min\{1,2m\exp(-c_{\mathrm{tail}}\Delta p)\},
 \qquad c_{\mathrm{tail}}\coloneqq8\log8-7.
\end{equation}

\begin{figure}[htbp]
 \centering
 \resizebox{\linewidth}{!}{\input{figures/star-conditional-fibre}}
 \caption{Resampling a star while keeping the outside edges fixed.
 Thin solid edges are fixed and present. Thick solid edges are present
 star edges; dotted edges are absent. The first two assignments are
 triangle-free. The third contains $uvw$, so it has zero probability
 under the conditional distribution on the star.}
 \label{fig:star-fibre}
\end{figure}
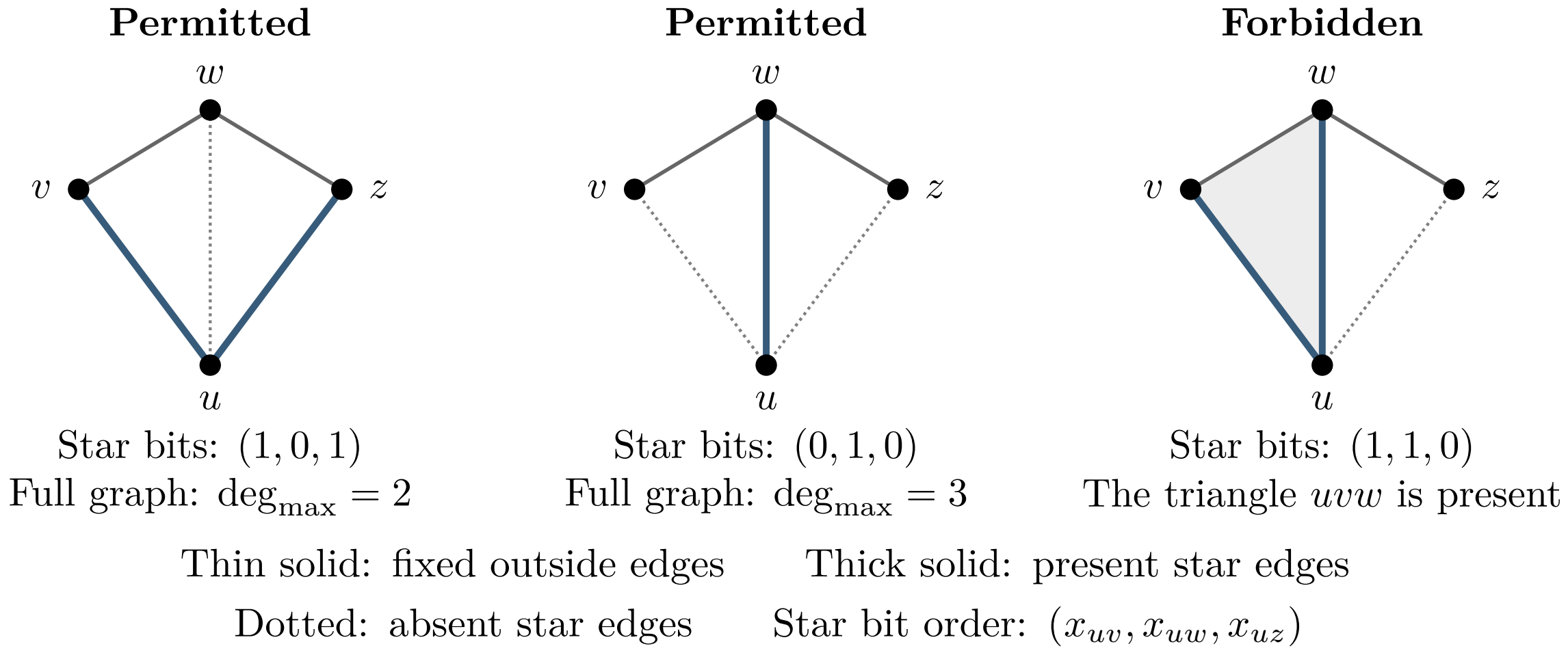

The bounds \eqref{eq:triangle-star-energy} and
\eqref{eq:triangle-difference-average} use different degree events. For
\cref{lem:common-set}, the good event must be determined outside the star,
and the interaction bound must hold for every permitted assignment within
it. In contrast, \cref{lem:difference-average} allows the good event
$\deg_{\max}\le d_0$ to depend on the full input.
\Cref{fig:star-fibre} illustrates the distinction. Take
$B\coloneqq\st(u)=\{uv,uw,uz\}$ and fix the two outside edges $vw,wz$
to be present, with no other possible edges. The interaction count for
$uw$ is $N_{uw,B}(\vx)=x_{vw}+x_{wz}=2$ for every permitted star assignment,
whether or not $uw$ is present. However, the first two permitted
assignments have maximum degrees two and three, respectively. Thus a
bound on the degree of the current full graph need not remain valid
under resampling within the star. We use the maximum degree of the
outside graph for \eqref{eq:triangle-star-energy}, and $\deg_{\max}$ for
\eqref{eq:triangle-difference-average}.

\begin{lemma}[Bounds for star averages]\label[lemma]{lem:triangle-star}
For every vertex $w$, every $R>0$, and every negative joint vector $\vf$,
\begin{equation}\label{eq:triangle-star-energy}
 \sum_{i\in\st(w)}\ip{\Deriv_i\vf}{\Filter_w\Deriv_i\vf}
 \le3R\norm\vf^2+\Delta b\norm\vf_\infty^2.
\end{equation}
Put $\tau^2\coloneqq144p^2\Delta^2b$. For every triangle $T$, coordinate
$i\in I$, and $R\ge1$,
\begin{equation}\label{eq:triangle-difference-average}
 \norm{\Deriv_i\Filter_T\vf}^2
 \le4\norm{\Filter_T\Deriv_i\vf}^2
       +\frac pR\norm\vf^2+\frac{4\tau^2}{R^2}\norm\vf_\infty^2.
\end{equation}
\end{lemma}
\begin{proof}
For \eqref{eq:triangle-star-energy}, fix all edges outside $\st(w)$. A
permitted set of present star edges cannot contain both $wu$ and $wv$ when the outside
edge $uv$ is present. These are exactly the restrictions: a triangle
through $w$ has two star edges and one outside edge, and the fixed outside
assignment already contains no triangle. Thus the conditional distribution on the star
is that of independent $\Bernoulli(p)$ bits conditioned on avoiding these forbidden
pairs. For $i=wu$, the count $N_{i,\st(w)}(\vx)$ is the number of present
outside edges $uv$ for which $wv\in\st(w)$. It depends only on the outside
assignment and is at most the degree of $u$ in that outside graph.
Thus an outside assignment with maximum degree at most $d_0$ gives
$N_{i,\st(w)}(\vx)\le d_0$ for every $i\in\st(w)$, uniformly over all
permitted assignments inside the star. The probability of a bad outside
assignment is at most $b$, because deleting a star cannot increase maximum
degree. This good event depends only on outside coordinates. By
\eqref{eq:triangle-cutoff}, \cref{lem:common-set} applies and gives
\eqref{eq:triangle-star-energy}.

For \eqref{eq:triangle-difference-average}, use the event $\deg_{\max}\le d_0$ in
\cref{lem:difference-average}. Fix $i=uv$. An edge that can complete a
triangle with $i$ is of the form $uw$ with $vw$ present, or $vw$ with $uw$
present. There are at most $\deg_u(\vx)+\deg_v(\vx)\le2d_0$ such edges at an
input $\vx$ in this event. Since each weight
$h_T(j)$ is at most two, their total weight is at most $4d_0$.
Using $\sum_jh_T(j)\le3\Delta$, the coefficient of
$\norm\vf_\infty^2$ in \eqref{eq:difference-average} is at most
$64p^2(3\Delta)^2b/R^2=4\tau^2/R^2$.
The numerical hypotheses also hold: $p(4d_0)\le36\Delta p^2\le0.36<2/5$.
Moreover, $\Delta p\ge1$ and $p\le1/(10\sqrt\Delta)$ imply
$\Delta\ge100$ and $p\le1/100$, whence
$4d_0\le36\Delta p\le3\Delta$.
Applying \cref{lem:difference-average} proves
\eqref{eq:triangle-difference-average}.
\end{proof}

\paragraph{Summing over triangles.}\label{sec:triangle-sums}
We now sum the conditional variances over the triangles that meet a queried
edge. By \eqref{eq:triangle-average-order}, terms from different triangles that
use the star at $w$ can all be bounded with $\Filter_w$. For an edge
$i\in T$ and either endpoint $w$ of $i$,
\begin{equation}\label{eq:triangle-shared-star}
 \norm{\Filter_T\Deriv_i\vf}^2
 =\ip{\Deriv_i\vf}{\Filter_T^2\Deriv_i\vf}
 \le\ip{\Deriv_i\vf}{\Filter_w\Deriv_i\vf}.
\end{equation}
When several triangle terms use the same edge $i$ and star at $w$, the term
$\ip{\Deriv_i\vf}{\Filter_w\Deriv_i\vf}$ is counted with that multiplicity.

Fix a queried edge $e=uv$ and define
\[
 \mathcal A_e\coloneqq\{T\in\Triangles:e\in T\},\qquad
 \mathcal B_{u,v}\coloneqq
       \{T\in\Triangles:u\in V(T),\ v\notin V(T)\}.
\]
Up to exchanging $u$ and $v$, these are the two ways that a triangle can
meet the queried edge: a triangle containing both endpoints contains $e$ itself.

\begin{lemma}[Two triangle sums]\label[lemma]{lem:triangle-sums}
For every negative joint vector $\vf$ and $R\ge1$,
\begin{align}
 \sum_{T\in\mathcal A_e}\sum_{i\in T\setminus\{e\}}
       \norm{\Deriv_i\Filter_T\vf}^2
 &\le c\left(R+\frac{\Delta p}R\right)\norm\vf^2
       +c\left(\Delta b+\frac{\Delta\tau^2}{R^2}\right)\norm\vf_\infty^2,
       \label{eq:triangle-sum-containing}\\
 \sum_{T\in\mathcal B_{u,v}}\sum_{i\in T}
       \norm{\Deriv_i\Filter_T\vf}^2
 &\le c\left(\Delta R+\frac{\Delta^2p}R\right)\norm\vf^2
       +c\left(\Delta^2b+\frac{\Delta^2\tau^2}{R^2}\right)\norm\vf_\infty^2.
       \label{eq:triangle-sum-meeting}
\end{align}
\end{lemma}
\begin{proof}
For \eqref{eq:triangle-sum-containing}, we first bound the sum of
$\norm{\Filter_T\Deriv_i\vf}^2$. For
$T=\{uv,uw,vw\}\in\mathcal A_e$, bound the term for $uw$ using the
star at $u$, and the term for $vw$ using the star at $v$. Each edge occurs
at most once in its respective star. By \eqref{eq:triangle-shared-star}
and \eqref{eq:triangle-star-energy},
\begin{equation}\label{eq:triangle-sum-right-containing}
 \sum_{T\in\mathcal A_e}\sum_{i\in T\setminus\{e\}}
       \norm{\Filter_T\Deriv_i\vf}^2
 \le c(R\norm\vf^2+\Delta b \norm\vf_\infty^2).
\end{equation}
There are at most $\Delta$ such triangles and hence at most $2\Delta$
terms. Applying \eqref{eq:triangle-difference-average} to each proves
\eqref{eq:triangle-sum-containing}.

For \eqref{eq:triangle-sum-meeting}, let
$T=\{uw,uz,wz\}\in\mathcal B_{u,v}$. The edges $uw,uz$ belong
to the star at $u$. Each fixed edge there belongs to at most $\Delta$
triangles. The terms for these edges sum to at most $\Delta$ times the
right side of \eqref{eq:triangle-star-energy}. For the opposite edge
$wz$, choose its first endpoint in a fixed vertex order and use that
endpoint's star. There are at most $\Delta$
choices of this endpoint, because it is adjacent to $u$ in $G$. For each
chosen endpoint, an edge $wz$ occurs for at most one triangle with the
fixed vertex $u$. Applying \eqref{eq:triangle-star-energy} to each such star gives
\begin{equation}\label{eq:triangle-sum-right-meeting}
 \sum_{T\in\mathcal B_{u,v}}\sum_{i\in T}
       \norm{\Filter_T\Deriv_i\vf}^2
 \le c(\Delta R \norm\vf^2+\Delta^2b \norm\vf_\infty^2).
\end{equation}
There are at most $\binom\Delta2$ triangles through $u$, so this sum has
at most $3\binom\Delta2$ terms. Another application of
\eqref{eq:triangle-difference-average} proves
\eqref{eq:triangle-sum-meeting}.
\end{proof}

\subsection{Query error and lower bound}\label{sec:triangle-query}
We now average first and apply the four-input comparison afterwards.
Define the comparison map $\Ups:\X\to\Y$ by
\begin{equation}\label{eq:full-transfer}
 \Ups\vf\coloneqq\frac1{\sqrt M}
       \sum_{T\in\Triangles}\sum_{\vy\in\supp\mu_T}
          \sqrt{\mu_T(\vy)}\,
          (\Local_T\Filter_T\vf)_{\vy}\otimes\ket{T,\vy}.
\end{equation}
Its component for fixed $T$ is $\Local_T\Filter_T\vf$, with the convention
in \eqref{eq:positive-space}. Since
$\Filter_T\ket{\mu_-}=\ket{\mu_-}$ and
$\Local_T\ket{\mu_-}=\ket{\mu_T}$, we have
$\Ups\ket{\mu_-}=\ket{\mu_+}$.

The query phases on the positive input and the four negative inputs need
not agree. The query also need not commute with the preceding average.
For the component of the error with fixed $T$, write
\begin{equation}\label{eq:triangle-query-split}
 \begin{aligned}
 \vg_T&\coloneqq
   (\Phase_e^+\Local_T\Filter_T-\Local_T\Filter_T\Phase_e^-)\vf\\
 &= (\Phase_e^+\Local_T-\Local_T\Phase_e^-)\Filter_T\vf
       +\Local_T[\Phase_e^-,\Filter_T]\vf.
 \end{aligned}
\end{equation}
By \eqref{eq:positive-space}, the squared query error is
$M^{-1}\sum_T\norm{\vg_T}^2$.

\begin{lemma}[One queried edge]\label[lemma]{lem:triangle-query}
For $R\ge1$, there are nonnegative numbers $\alpha,\beta$ such that,
uniformly in $e\in I$ and the workspace,
\begin{equation}\label{eq:triangle-query-bound}
 \norm{(\Phase_e^+\Ups-\Ups\Phase_e^-)\vf}^2
       \le\alpha\norm\vf^2+\beta\norm\vf_\infty^2,
\end{equation}
where
\begin{align}
 \alpha&\le\frac cM\left(
       \frac Rp+\frac\Delta R+\frac\Delta{R^2}
                    +\frac{\Delta^2p}{R^2}\right),
       \label{eq:triangle-alpha}\\
 \beta&\le\frac cM\left(
       \frac{\Delta b}p+\frac{\Delta\tau^2}{pR^2}
       +\frac{\Delta^2b}{R^2}+\frac{\Delta^2\tau^2}{pR^4}\right).
       \label{eq:triangle-beta}
\end{align}
\end{lemma}
\begin{proof}
Recall the commutator $\Comm_e=[\Deriv_e,\Phase_e^-]$ from
\eqref{eq:query-commutator}, here under the negative distribution. By
\eqref{eq:same-coordinate},
$\norm{\Comm_e}\le2\sqrt p$ and
$\norm{\Comm_e\vf}_\infty\le2\norm\vf_\infty$.
For $i\ne e$, $[\Deriv_i,\Phase_e^-]=0$, so
\eqref{eq:filter-Q} specializes to
\begin{equation}\label{eq:triangle-phase-average}
 [\Phase_e^-,\Filter_T]
       =\frac{h_T(e)}R\Filter_T\Comm_e\Filter_T,
 \qquad
 \norm{[\Phase_e^-,\Filter_T]}\le\frac{2h_T(e)\sqrt p}R.
\end{equation}
If $T$ contains neither endpoint of $e$, both terms in
\eqref{eq:triangle-query-split} vanish.

Suppose first that $T\in\mathcal A_e$. By
\eqref{eq:transfer-defect} and \eqref{eq:triangle-sum-containing}, the
sum of the squared first terms in \eqref{eq:triangle-query-split} is at
most
\[
 c\left(\frac Rp+\frac\Delta R\right)\norm\vf^2
   +c\left(\frac{\Delta b}p
                    +\frac{\Delta\tau^2}{pR^2}\right)\norm\vf_\infty^2.
\]
For the second term, \eqref{eq:transfer-rough} and
\eqref{eq:triangle-phase-average} give
\[
 \norm{\Local_T[\Phase_e^-,\Filter_T]\vf}^2
 \le\frac cp\norm{[\Phase_e^-,\Filter_T]\vf}^2
 \le\frac{c\norm\vf^2}{R^2}
\]
for each triangle, since $h_T(e)=2$. Summing over the at most $\Delta$
candidates and applying Cauchy--Schwarz to the two terms in
\eqref{eq:triangle-query-split} gives
\begin{equation}\label{eq:triangle-query-containing}
 \sum_{T\in\mathcal A_e}\norm{\vg_T}^2
 \le c\left(\frac Rp+\frac\Delta R+\frac\Delta{R^2}\right)\norm\vf^2
       +c\left(\frac{\Delta b}p
                    +\frac{\Delta\tau^2}{pR^2}\right)\norm\vf_\infty^2.
\end{equation}

Next consider $T\in\mathcal B_{u,v}$. Here $e\notin T$ and
$h_T(e)=1$, so only the second term remains. Put
\[
 \boldsymbol q_T\coloneqq[\Phase_e^-,\Filter_T]\vf
       =R^{-1}\Filter_T\Comm_e\Filter_T\vf.
\]
There are at most $\binom\Delta2$ such triangles, and thus
\begin{equation}\label{eq:triangle-query-meeting-norm}
 \sum_{T\in\mathcal B_{u,v}}\norm{\boldsymbol q_T}^2
       \le c\frac{\Delta^2p}{R^2}\norm\vf^2.
\end{equation}
Using \eqref{eq:transfer-rough} here would lose the factor $p$ in
\eqref{eq:triangle-query-meeting-norm}. We instead use
\eqref{eq:transfer-energy}, which gives
\[
 \sum_{T\in\mathcal B_{u,v}}\norm{\Local_T\boldsymbol q_T}^2
 \le c\sum_{T\in\mathcal B_{u,v}}\norm{\boldsymbol q_T}^2
   +\frac cp\sum_{T\in\mathcal B_{u,v}}\sum_{i\in T}
             \norm{\Deriv_i\boldsymbol q_T}^2.
\]
It remains to bound the last sum.

For $i\in T$, we have $i\ne e$ and $[\Deriv_i,\Phase_e^-]=0$.
The Jacobi identity and \eqref{eq:two-coordinate} give
\begin{equation}\label{eq:triangle-query-difference-commutator}
 [\Deriv_i,\Comm_e]=[[\Deriv_i,\Deriv_e],\Phase_e^-],
 \qquad \norm{[\Deriv_i,\Comm_e]}\le2p.
\end{equation}
Apply \eqref{eq:triangle-difference-average} to
$\vh\coloneqq\Comm_e\Filter_T\vf$ and divide by $R^2$. The bounds
$\norm\vh\le2\sqrt p\,\norm\vf$, $\norm\vh_\infty\le2\norm\vf_\infty$, and
\[
 \Deriv_i\vh=\Comm_e\Deriv_i\Filter_T\vf
               +[\Deriv_i,\Comm_e]\Filter_T\vf
\]
yield
\begin{equation}\label{eq:triangle-query-difference}
 \begin{aligned}
 \norm{\Deriv_i\boldsymbol q_T}^2
 &\le\frac4{R^2}\norm{\Filter_T\Deriv_i\vh}^2
       +\frac p{R^3}\norm\vh^2
       +\frac{4\tau^2}{R^4}\norm\vh_\infty^2\\
 &\le c\frac p{R^2}\norm{\Deriv_i\Filter_T\vf}^2
       +c\frac{p^2}{R^2}\norm\vf^2
       +c\frac{\tau^2}{R^4}\norm\vf_\infty^2.
 \end{aligned}
\end{equation}
The last inequality uses $R\ge1$ and that $\Filter_T$ is a contraction.
Now sum over $T\in\mathcal B_{u,v}$ and $i\in T$, and apply
\eqref{eq:triangle-sum-meeting}. There are $O(\Delta^2)$ pairs; using
$R\ge1$ and $p\le1$ gives
\begin{equation}\label{eq:triangle-query-difference-sum}
 \sum_{T\in\mathcal B_{u,v}}\sum_{i\in T}
          \norm{\Deriv_i\boldsymbol q_T}^2
 \le c\left(\frac{\Delta p}R
                    +\frac{\Delta^2p^2}{R^2}\right)\norm\vf^2
       +c\left(\frac{\Delta^2pb}{R^2}
                    +\frac{\Delta^2\tau^2}{R^4}\right)\norm\vf_\infty^2.
\end{equation}
Combining \eqref{eq:transfer-energy},
\eqref{eq:triangle-query-meeting-norm}, and
\eqref{eq:triangle-query-difference-sum}, we obtain
\begin{equation}\label{eq:triangle-query-meeting}
 \sum_{T\in\mathcal B_{u,v}}\norm{\Local_T\boldsymbol q_T}^2
 \le c\left(\frac{\Delta^2p}{R^2}+\frac\Delta R\right)\norm\vf^2
       +c\left(\frac{\Delta^2b}{R^2}
                    +\frac{\Delta^2\tau^2}{pR^4}\right)\norm\vf_\infty^2.
\end{equation}
Equation~\eqref{eq:triangle-query-meeting} also holds for $\mathcal B_{v,u}$.
Adding its two instances and \eqref{eq:triangle-query-containing}, then dividing by $M$ as in
\eqref{eq:positive-space}, proves
\eqref{eq:triangle-query-bound}--\eqref{eq:triangle-beta}.
\end{proof}

\paragraph{Choosing the amount of averaging.}
To balance $R/p$ and $\Delta^2p/R^2$ in \eqref{eq:triangle-alpha}, take
\begin{equation}\label{eq:triangle-R}
 R\coloneqq(\Delta p)^{2/3}.
\end{equation}
Since $\Delta p\ge1$, we have $R\ge1$ and the other terms in
\eqref{eq:triangle-alpha} are no larger. Thus
\begin{equation}\label{eq:triangle-ordinary}
 \alpha\le c\frac{\Delta^{2/3}}{Mp^{1/3}}.
\end{equation}

\begin{proof}[Proof of \cref{thm:fixed-graph}]
Use the comparison \eqref{eq:full-transfer} with the choice
\eqref{eq:triangle-R}. Put
$A_0\coloneqq\Delta^{2/3}/(Mp^{1/3})$. We have $\alpha\le cA_0$ by
\eqref{eq:triangle-ordinary}. To apply \cref{lem:transfer}, it remains to
show $m\beta\le cA_0$. Substituting
$\tau^2=144p^2\Delta^2b$ and \eqref{eq:triangle-R} into
\eqref{eq:triangle-beta}, the four terms divided by $A_0$ are at most
absolute constants times
\[
 b\Delta^{1/3}p^{-2/3},\qquad b\Delta,\qquad
 b/p,\qquad b\Delta^{2/3}p^{-4/3},
\]
respectively. Since $p\ge1/\Delta$, each is at most $cb\Delta^2$.
There are $m$ query indices, so
\begin{equation}\label{eq:triangle-exceptional}
 \frac{m\beta}{A_0}\le c m b\Delta^2
       \le c m^4\exp(-c_{\mathrm{tail}}\Delta p),
\end{equation}
where we used $\Delta\le m$ and \eqref{eq:triangle-tail}.
Taking $C_0$ sufficiently large in \eqref{eq:density-range} makes the
last expression bounded by an absolute constant, uniformly in $m$.

To finish, an output consisting of three distinct vertices is checked
for membership in $\Triangles$ without queries, and three queries verify its
edges. This verifier always rejects on $\mu_-$. By \cref{lem:transfer},
\eqref{eq:triangle-ordinary}, and \eqref{eq:triangle-exceptional},
\[
 \sqrt\sigma\le(t+3)\sqrt{\alpha+m\beta}
       \le C_1(t+3)\frac{\Delta^{1/3}}{\sqrt M\,p^{1/6}}.\qedhere
\]
\end{proof}

\begin{proof}[Proof of \cref{thm:main-result}]
Take $G$ to be the complete tripartite graph $K_{n,n,n}$, whose three parts
have size $n$ and are known to the algorithm. Every input graph has edges
only between different parts. Then $m=3n^2$, $\Delta=2n$, and $M=n^3$.
Set $p\coloneqq1/(10\sqrt{2n})$. The first condition in
\eqref{eq:density-range} holds with equality, and for sufficiently large $n$,
\[
 \Delta p=\frac{\sqrt{2n}}{10}\ge C_0\log(6n^2).
\]
Thus \cref{thm:fixed-graph} gives
\[
 \sqrt\sigma\le C_1(t+3)\frac{(2n)^{1/3}}{n^{3/2}p^{1/6}}
       \le c(t+3)n^{-13/12}.
\]
An algorithm succeeding with probability at
least $2/3$ on every promised positive input has at least this average
success, and therefore needs $\Omega(n^{13/12})$ queries.
For an arbitrary number $N$ of vertices, take
$n\coloneqq\lfloor N/3\rfloor$ and leave the remaining vertices isolated.
All additional pairs have known answer zero. This proves the
$\Omega(N^{13/12})$ lower bound.
\end{proof}

\begin{nonbipartitecorollary}
For every fixed connected non-bipartite graph $H$, detecting $H$ as a
subgraph of an $N$-vertex graph in the adjacency matrix model requires
$\Omega(N^{13/12})$ quantum queries.
\end{nonbipartitecorollary}

\begin{proof}[Proof of \cref{cor:non-bipartite}]
By~\cite[Corollary 5.3]{GZ26}, detecting a fixed connected non-bipartite
graph $H$ is at least as hard as detecting a triangle, up to constant
factors in the number of vertices and queries. The result follows from
\cref{thm:main-result} and the constant-factor search-to-decision
reduction for triangles~\cite[Lemma 23]{CGP26}.
\end{proof}

%% file: figures/star-conditional-fibre.tex
\begingroup%
\colorlet{ink}{black}%
\definecolor{accent}{RGB}{55,91,122}%
\colorlet{context}{black!60}%
\colorlet{absent}{black!50}%
\begin{tikzpicture}[x=1cm,y=1cm,
  star/.style={line width=1.7pt,accent},
  outside/.style={line width=.95pt,context},
  missing/.style={densely dotted,line width=.75pt,absent},
  vertex/.style={circle,fill=ink,inner sep=1.9pt},
  every node/.style={text=ink,font=\small}]
\foreach \j in {0,1,2}{
  \begin{scope}[xshift=\j*4.9cm]
    \coordinate (u) at (0,0);
    \coordinate (v) at (-1.16,1.55);
    \coordinate (w) at (0,2.25);
    \coordinate (z) at (1.16,1.55);
    \ifnum\j=2
      \fill[black!7] (u)--(v)--(w)--cycle;
    \fi
    \draw[missing] (u)--(v) (u)--(w) (u)--(z);
    \draw[outside] (v)--(w)--(z);
    \ifnum\j=0 \draw[star] (v)--(u)--(z);\fi
    \ifnum\j=1 \draw[star] (u)--(w);\fi
    \ifnum\j=2 \draw[star] (v)--(u)--(w);\fi
    \node[vertex,label=below:$u$] at (u) {};
    \node[vertex,label=left:$v$] at (v) {};
    \node[vertex,label=above:$w$] at (w) {};
    \node[vertex,label=right:$z$] at (z) {};
    \ifnum\j<2
      \node[font=\small\bfseries] at (0,3.03) {Permitted};
    \else
      \node[font=\small\bfseries] at (0,3.03) {Forbidden};
    \fi
    \ifnum\j=0
      \node at (0,-.73) {Star bits: $(1,0,1)$};
      \node at (0,-1.16) {Full graph: $\deg_{\max}=2$};
    \fi
    \ifnum\j=1
      \node at (0,-.73) {Star bits: $(0,1,0)$};
      \node at (0,-1.16) {Full graph: $\deg_{\max}=3$};
    \fi
    \ifnum\j=2
      \node at (0,-.73) {Star bits: $(1,1,0)$};
      \node at (0,-1.16) {The triangle $uvw$ is present};
    \fi
  \end{scope}
}
\node[align=center,text width=14.3cm] at (4.9,-2.05)
  {Thin solid: fixed outside edges\qquad
   Thick solid: present star edges\\[3pt]
   Dotted: absent star edges\qquad
   Star bit order: $(x_{uv},x_{uw},x_{uz})$};
\path[use as bounding box]
  ([xshift=-7pt,yshift=-7pt]current bounding box.south west)
  rectangle ([xshift=7pt,yshift=7pt]current bounding box.north east);
\end{tikzpicture}%
\endgroup